\documentclass[aps, prx, twocolumn, longbibliography, superscriptaddress]{revtex4-1}
\usepackage{bm, amsmath, amsfonts, amssymb, braket,mathtools,upgreek,amsthm}
\DeclareMathOperator{\Tr}{Tr}

\usepackage{multirow}
\usepackage{float}
\usepackage{color}
\usepackage{comment}
\usepackage{subfigure}
\usepackage{lipsum}
\usepackage[colorlinks,urlcolor=blue,citecolor=blue,linkcolor=black]{hyperref}

\begin{document}

\title{Non-Hermitian bulk-boundary correspondence via scattering theory}

\author{Haoshu Li}
\email{lihaoshu@mail.ustc.edu.cn}
\affiliation{Department of Physics, University of Science and Technology of China, Hefei 230026, China}

\author{Qian Niu}
\affiliation{ICQD and Department of Physics, University of Science and Technology of China, Hefei 230026, China}

\begin{abstract}
    The conventional bulk-boundary correspondence breaks down in non-Hermitian systems. In this paper, we reestablish the bulk-boundary correspondence in one-dimensional non-Hermitian systems by applying the scattering theory, which is a systematical way in various symmetry classes. Based on the scattering theory, it is discovered that the topological invariant can be obtained by solving a generalized eigenproblem without calculating the generalized Brillouin zone. As a direct consequence, we unveil a new type of topological phase transition without typical bulk enengy gap closing and an unstable phase with topological boundary states, dubbed the critical topological phase.
\end{abstract}

\maketitle

\section{Introduction}
In topological insulators and topological superconductors  \cite{PhysRevLett.49.405, PhysRevB.41.9377, RevModPhys.82.3045, science.1133734, PhysRevLett.95.226801, PhysRevLett.95.146802, PhysRevB.74.195312, science.1148047, PhysRevB.76.045302, PhysRevLett.98.106803, PhysRevB.78.195424, RevModPhys.83.1057, RevModPhys.88.035005}, there are quantized quantities so-called topological invariants which are robust under perturbation without closing the bulk energy gap. Topological invariants protected by the internal symmetries can be $\mathbb{Z}$ and $\mathbb{Z}_2$ valued. In topological insulators and topological superconductors, the bulk-boundary correspondence (BBC) which relates the quantized topological invariants to the protected gapless excitations localized at the boundary is of great interests. In Hermitian systems, there are different approaches to establish the BBC including the index theorem \cite{blms.5.2.229, PhysRevD.24.2669, RevModPhys.88.035001, PhysRevLett.128.251601}, the Green's function \cite{PhysRevB.84.125132}, the Wilson loop \cite{PhysRevLett.107.036601}, and the scattering matrix \cite{PhysRevB.83.155429, PhysRevB.85.165409, PhysRevB.95.235143}.

The conventional BBC of Hermitian systems relates the topological invariants defined for infinite periodic systems to the protected gapless boundary excitations occurring in systems with open boundaries. In recent studies of non-Hermitian topological phases, the BBC in its familiar form is found to generically break down due to the appearance of the localized bulk states, which is called the non-Hermitian skin effect (NHSE) \cite{yao2018, yao20182, RevModPhys.93.015005}. Although bulk states may be localized at the boundary, boundary states and bulk states can still be distinguished by their energies since energies of boundary states belong to the discrete spectrum while energies of the bulk states belong to the continuum spectrum. Several amendments to reestablish a modified BBC in the $\mathbb{Z}$ classified non-Hermitian topological phase have been proposed \cite{PhysRevLett.121.026808, Xiong_2018, yao2018, yao20182, PhysRevLett.125.226402, RevModPhys.93.015005, PhysRevLett.123.246801, PhysRevB.99.201103}. On the other hand, the BBC of $\mathbb{Z}_2$ non-Hermitian topological phases are rarely studied in the literature. Hence, we want a method to be able to describe the modified BBC in $\mathbb{Z}$ and $\mathbb{Z}_2$ non-Hermitian topological phases in the same framework.

An important approach to reestablish a modified BBC in the one-dimensional system is the generalized Brillouin zone (GBZ) method \cite{yao2018, PhysRevLett.123.066404}. 
The GBZ is the momentum spectrum of bulk states on the complex plane as a generalization of the Brillouin zone in non-Hermitian systems.
Instead of calculating the topological invariant by integration on the Brillouin zone, the topological invariant is obtained by integration on the GBZ in the modified BBC relation \cite{yao2018, yao20182, PhysRevLett.125.226402}. Although the GBZ approach of the modified BBC is a great way to gain understanding, in practice, calculating the GBZ can be complicated and time-consuming. Especially, for systems with multi-subGBZs, calculating sub-GBZs is cumbersome \cite{PhysRevLett.125.226402}. This is another reason for us to develop an alternative approach to calculate the topological invariant that defines the modified BBC. 

The key of establishing the modified BBC is to find the correct expression of the topological invariant. In Hermitian systems, one of ways to express the topological invariant is using the reflection matrix \cite{PhysRevB.83.155429, PhysRevB.85.165409, PhysRevB.95.235143}. We discover that in non-Hermitian systems, the topological invariant in terms of the reflection matrix is still the correct topological invariant to establish the BBC. This method overcomes disadvantages of the GBZ approach discussed above and works for all symmetry classes. Based on this reflection matrix method, the topological invariant can be calculated analytically by solving a generalized eigenproblem without computing the GBZ. Interestingly, we find that some non-Hermitian systems with discrete symmetries or constraints go beyond Hermitian results, which implies a new type of topological phase transition without typical bulk enengy gap closing and an unstable phase with topological boundary states.

The paper is organized as following. In Sec.~\ref{sec: scattering}, we introduce the scattering matrix method used in this paper. In Sec.~\ref{sec: SLS}, we derive a quantized topological invariant in non-Hermitian systems with the sublattice symmetry via the scattering matrix method: we establish the BBC in Sec.~\ref{sec: BBC_SLS}, give an analytical method based on solving a generalized eigenproblem to calculate the topological invariant in Sec.~\ref{sec: analy} and show a critical topological phase in Sec.~\ref{sec: critical_topo}.
In Sec.~\ref{sec: TR}, we derive the $\mathbb{Z}_2$ topological invariant and establish the BBC for non-Hermitian systems with time-reversal and sublattice symmetries. In Sec.~\ref{sec: Conclusions}, we present our conclusions.

\section{Scattering theory}
\label{sec: scattering}
In photonics, non-Hermitian physics can be described by an analogy of the Schr{\"o}dinger equation \cite{Regensburger2012, Rechtsman2013, Weimann2017, doi:10.1080/00018732.2021.1876991} as well as the scattering formalisim \cite{RevModPhys.69.731, doi:10.1080/00018732.2021.1876991}. 

\begin{figure}
    \includegraphics[width=0.9\columnwidth]{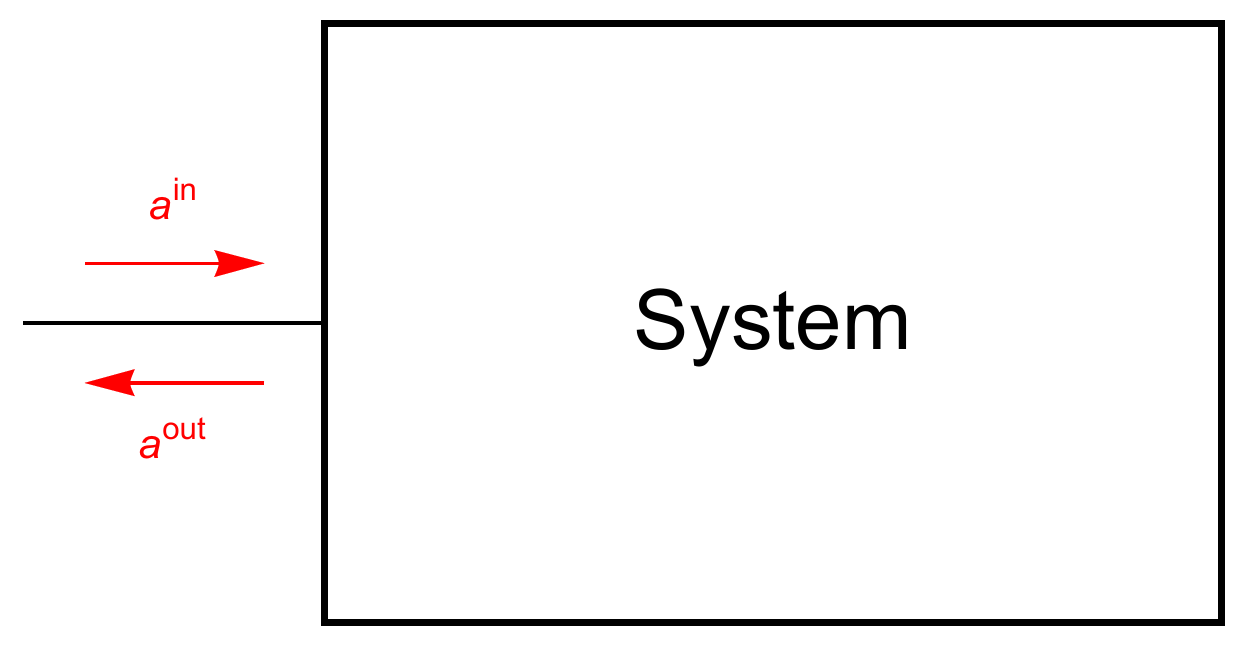}
    \caption{Schematic figure illustrating the input-output scattering formalism for the left channel case. The scattering medium has a left optical channel and $\bm{a}^{\text{in(out)}}$ represents the the electromagnetic fields of the incoming (outgoing) modes at the left channel.}
    \label{fig: scattering}
\end{figure}

In this paper, we study one-dimensional translationally invariant non-Hermitian systems. To obtain the topological properties of the system, we attach the system with a left optical channel as shown in Fig.~\ref{fig: scattering}. Consider the optical amplitudes with frequency $\omega$ at the left optical channel, they are related by the reflection matrix $r(\omega)$ as $a^{\text{out}} = r(\omega) a^{\text{in}}$.

By the Mahaux-Weidenm{\"u}ller formula (see Appendix \ref{sec: Mahaux} for detail), the reflection matrix $r$ from the scattering matrix $S$ can be expressed in terms of the Green's function of the system as 
\begin{align}
    r(\omega) = \frac{\mathbb{I} + i V_{\text{LS}} G(\omega) V^{\dagger}_{\text{LS}}}{\mathbb{I} - i V_{\text{LS}} G(\omega) V^{\dagger}_{\text{LS}}},
    \label{eq: Mahaux}
\end{align}
where $V_{\text{LS}}$ is the coupling between the system and the left channel and $G(\omega)=\frac{1}{\omega-H}$ is the real space Green's function of the system with the Hamiltonian $H$. Assume that the coupling between the system and the left channel is only at the first unit cell of the system, the reflection matrix $r$ can be further expressed as 
\begin{align}
    r(\omega) = \frac{\mathbb{I} + i V_{\text{LS}} G_{11}(\omega) V^{\dagger}_{\text{LS}}}{\mathbb{I} - i V_{\text{LS}} G_{11}(\omega) V^{\dagger}_{\text{LS}}},
    \label{eq: r_G}
\end{align}
where $G_{11}(\omega)$ is the diagonal block of the real space Green's function at the first unit cell from the left to the right.

In the following sections, we will generalize the result of Ref.~\cite{PhysRevB.83.155429} to non-Hermitian systems that expressing the topological invariants protected by different symmetries via the reflection matrix $r(\omega)$ and establish the BBC of non-Hermitian systems. This approach has the advantage that the topological invariants in terms of the reflection matrix $r(\omega)$ is measurable due to the fact that the reflection matrix $r(\omega)$ can be measured in experiments, such as in coupled-resonators systems \cite{peng2014loss, Longhi2015, doi:10.1126/science.aay1064}. 

Since there is no topological phase without any discrete symmetry in one-dimensional systems \cite{RevModPhys.88.035005}, the simplest topological phase in one-dimensional systems are protected by the sublattice symmetry, and we will discuss it in the next section.

\section{Systems with sublattice symmetry}
\label{sec: SLS}
Non-Hermitian system with the sublattice symmetry satisfies 
\begin{align}
    \Gamma H \Gamma^{-1} = -H,
\end{align}
where $\Gamma$ denotes the sublattice symmetry operator and is a unitary matrix, and $H$ is the real space Hamiltonian of the system. 

It is first discovered that in the non-Hermitian Su-Schrieffer-Heeger model with the sublattice symmetry, the conventional BBC breaks down. It is due to the occurrence of the NHSE as revealed in Ref.~\cite{yao2018}, i.e., the majority number of bulk eigenstates of the non-Hermitian systems may be localized at the boundary instead of being Bloch waves as in Hermitian systems. Hence, the attempt that using the Bloch Hamiltonian directly to characterize the BBC fails in non-Hermitian systems. In the reestablishment of the BBC in non-Hermitian systems, the GBZ is often used to calculate the topological invariant \cite{yao2018, yao20182, PhysRevLett.125.226402}. Here, we adopt a different approach that expressing the topological invariant in terms of the reflection matrix of the system. Hence, in the following, we attach the system with a fictitious left optical channel (or a fictitious left lead) to obatin the topological invariants.

Due to the sublattice symmetry, the real space Green's function satisfies 
\begin{align}
    \Gamma \, G(\omega) \, \Gamma^{-1} = -G(-\omega),
\end{align}
where $G(\omega) = \frac{1}{\omega - H}$ is the real space Green's function.

We consider the case that 
\begin{align}
    [V_{\text{LS}}, \Gamma] & = 0 = [V^{\dagger}_{\text{LS}}, \Gamma] \; \; \text{or} \notag \\
    \{V_{\text{LS}}, \Gamma\} & = 0 = \{V^{\dagger}_{\text{LS}}, \Gamma\},
\end{align}
i.e., the coupling between the system and the left channel preserves the sublattice symmetry.
It follows that 
\begin{align}
    \Gamma \, r(\omega) \, \Gamma^{-1} = r(-\omega)^{-1},
\end{align}
which further implies that
\begin{align}
    \left[\lim_{\omega \rightarrow 0} \Gamma \, r(\omega) \right]^2 = \mathbb{I}.
\end{align}
Hence, the eigenvalues of $\lim_{\omega \rightarrow 0} \Gamma r(\omega)$ is quantized to be $\pm 1$. The number of positive (negative) eigenvalues of $\lim_{\omega \rightarrow 0} \Gamma r(\omega)$ is a topological invariant and is unchanged provided that the bulk energy gap of the system remains unclosed since the change of the number of positive (negative) eigenvalues of $\lim_{\omega \rightarrow 0} \Gamma r(\omega)$ requires $\lim_{\omega \rightarrow 0} r(\omega)$ to be singular which represents the occurrence of perfectly transmitting modes. Therefore, $\lim_{\omega \rightarrow 0} \frac{\Tr (\Gamma r(\omega))}{2} \in \mathbb{Z}$ characterizing the difference between the number of positive eigenvalues and the number of negative eigenvalues of $\lim_{\omega \rightarrow 0} \Gamma r(\omega)$ is a well-defined topological invariant protected by the sublattice symmetry.

Before the end of this discussion, we want to make the following comment. Although the eigenvalues of $\lim_{\omega \rightarrow 0} \Gamma r(\omega)$ is quantized to be $\pm 1$, eigenvalues of $\lim_{\omega \rightarrow 0} r(\omega)$ do not have unitary magnitudes in general. Since unlike the Hermitian cases, the reflection matrix $r$ is not necessarily unitary \footnote{Another difference compared to the Hermitian cases is $\lim_{\omega \rightarrow 0} \Gamma r(\omega)$ is not necessarily Hermitian \cite{PhysRevB.85.165409}}.

\subsection{Establish the BBC}
\label{sec: BBC_SLS}
By the spectral decomposition of the real space Green's function, in the thermodynamics limit,
\begin{align}
    G(\omega) & = \sum_n \ket{n R} \frac{1}{\omega-E_n} \bra{n L} \notag \\
    & = \sum_{E_n = 0} \ket{n R} \frac{1}{\omega} \bra{n L} + \sum_{E_n \neq 0} \ket{n R} \frac{1}{\omega - E_n} \bra{n L}.
\end{align}
The coefficient matrix of the $\frac{1}{\omega}$ term of $G_{11}(\omega)$ is 
\begin{align}
    A = \sum_{E_n = 0} \braket{1 | n R}\braket{n L| 1},
\end{align}
where $\braket{1 | n R}=\ket{n R(1)}$ is a column vector with $M$ components representing the amplitudes of the right eigenstate $\ket{n R}$ at the first unit cell and $\braket{n L| 1}=\bra{n L(1)}$ is a row vector with $M$ components representing the amplitudes of the left eigenstate $\bra{n L}$ at the first unit cell.

By Eq.~(\ref{eq: Topo_A}) in Appendix \ref{sec: Topo_inv_G}, we obtain the BBC in the thermodynamics limit 
\begin{align}
    \lim_{\omega \rightarrow 0} -\frac{\Tr (\Gamma r(\omega))}{2}  = \text{rank}(A \, \Pi_+) - \text{rank}(A \, \Pi_-),
    \label{eq: SLS_BBC}
\end{align}
where $A = \sum_{E_n = 0} \braket{1 | n R}\braket{n L| 1}$ characterizes all zero energy modes ``localized'' at the left end, here, the ``localization'' means that $\braket{1 | n R}\braket{n L| 1}$ has nonzero eigenvalues. This BBC is understood as followings, the left hand side of Eq.~(\ref{eq: SLS_BBC}) represents the bulk topological invariant and the right hand side of Eq.~(\ref{eq: SLS_BBC}) encodes the information about the number of zero energy modes ``localized'' at the left end. 

For Hermitian cases, since right eigenstates and left eigenstates are identical, the right hand side of Eq.~(\ref{eq: SLS_BBC}) is the difference between the number of right eigenstates with positive chirality and negative chirality in all right eigenstates with zero energy localizing at the left end. In contrast to the BBC in Hermitian cases that edge state localization is only about right eigenstates, edge state localization for non-Hermitian cases is about the multiplication of the amplitude of the left eigenstate and the right eigenstate. As a result, left end ``localization'' can be contributed by left eigenstates localized at the left end, which implies that some right eigenstates are localized at the right end \cite{yao2018}.

\begin{figure} 
    \centering
    \includegraphics[width=0.9\columnwidth]{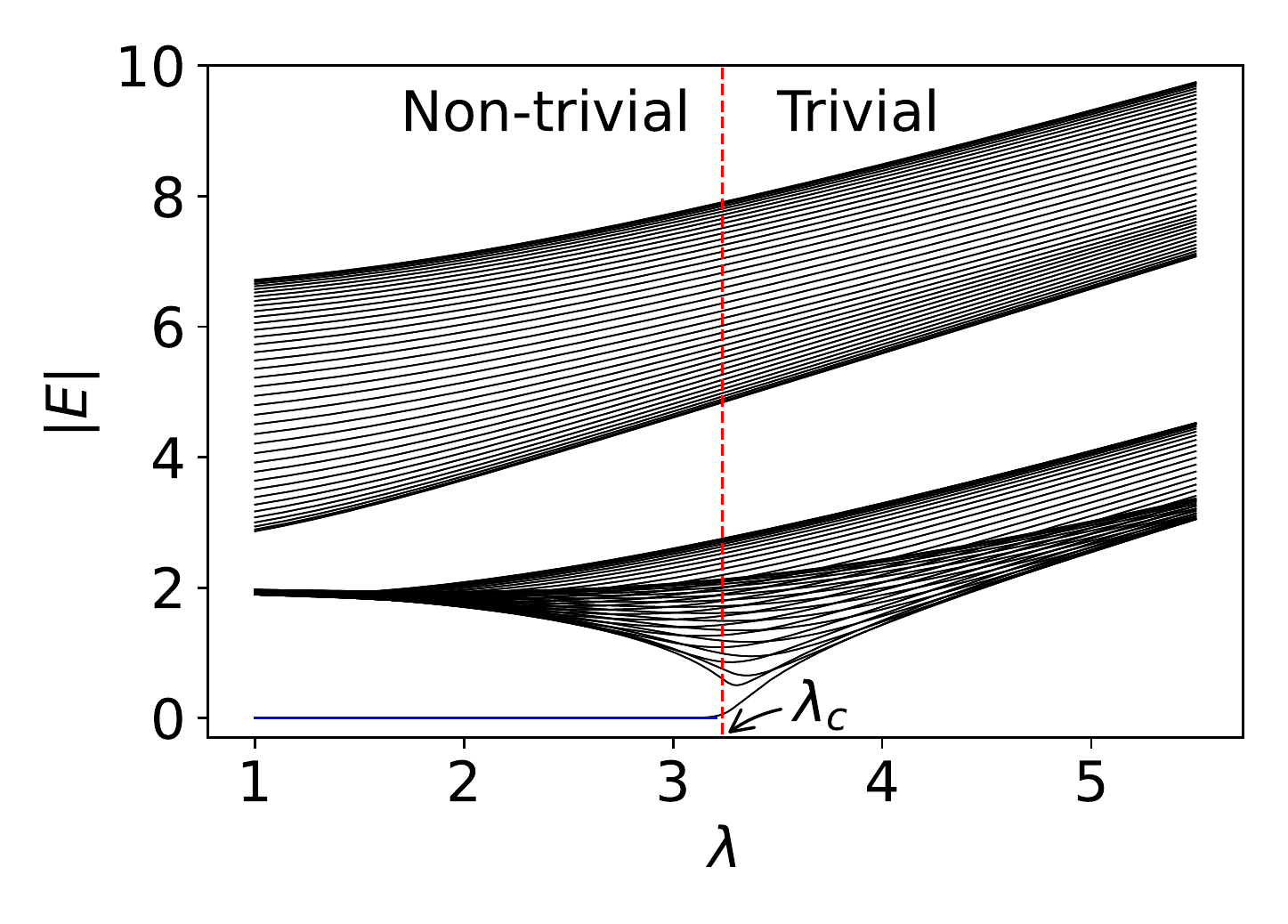}
    \caption[twoSubGBZs]{The energy spectrum (under the OBC) and topological invariant $-\frac{\Tr (\Gamma r(0))}{2}$ of the model vary with parameter $\lambda$. When $\lambda<\lambda_c$, this model is in a topological non-trivial phase with $-\frac{\Tr (\Gamma r(0))}{2}=1$. When $\lambda>\lambda_c$, this model is in a topological trivial phase with $-\frac{\Tr (\Gamma r(0))}{2}=0$. Blue line represents topological boundary states with discrete spectrum.}
    \label{fig: twoSubGBZs}
\end{figure}

As an example, we consider the following four-band model with the sublattice symmetry \cite{PhysRevLett.125.226402}, its Bloch Hamiltonian is
\begin{align}
    H(k) = \begin{pmatrix}
        0 & R_{+}(\beta) \\
        R_{-}(\beta) & 0
    \end{pmatrix},
\end{align}
where $R_{\pm}(\beta) = \lambda \mathbb{I}_{2 \times 2} + \frac{1}{2}(t_{\pm}+t_1 \beta^{\pm 1}) \sigma_{\pm} + \frac{1}{2}t_2 \beta^{\pm 1} \sigma_{\mp}$, $\beta=e^{ik}$, $\sigma_{\pm}=\sigma_x \pm i\sigma_y$ and $t_1=2,t_2=2i,t_{\pm}=5\pm 2$. This model has two non-degenerate sub-GBZs, the topological invariant of this model can be calculated by counting the number of zeros and poles of $R_{+}(\beta)$ and $R_{-}(\beta)$ encircled by each sub-GBZ. Compare to the GBZ method, the numerical scattering matrix method has the advantage that it does not involve the complexity of calculating two sub-GBZs and counting the number of zeros and poles encircled by each sub-GBZ. Futhermore, the numerical algorithm of the scattering matrix is more stable than any algorithm involving the spectral property manipulation since the computation of the scattering matrix only requires the matrix inversion which is more stable than the eigenvalue algorithm for non-Hermitian matrices. As shown in Fig.~\ref{fig: twoSubGBZs}, we plot the energy spectrum and the topological invariant $-\frac{\Tr (\Gamma r(0))}{2}$ for different values of $\lambda$, and the bulk-boundary correspondence is shown. 

\subsection{Analytical computation of the topological invariant}
\label{sec: analy}
The topological invariant is encoded in the coefficient matrix of the $\frac{1}{\omega}$ term of $G_{11}(\omega)$ block of the Green's function (see Appendix \ref{sec: Topo_inv_G}). In this section, we discuss the analytical method to calculate the coefficient matrix of the $\frac{1}{\omega}$ term of $G_{11}(\omega)$ and further the topological invariants in non-Hermitian systems with the sublattice symmetry. 

Without loss of generality, we consider the following tight-binding model with only the nearest hopping (one can always express the Hamiltonian with only the nearest hopping by extending the unit cell),
\begin{align}
    H = \sum_{n,\mu,\nu} c^{\dagger}_{n,\mu} h_{\mu,\nu} c_{n,\nu} + c^{\dagger}_{n+1,\mu} V_{\mu,\nu} c_{n,\nu} + c^{\dagger}_{n,\mu} W_{\mu,\nu} c_{n+1,\nu},
    \label{eq: main_Ham}
\end{align}
where $c^{\dagger}_{n,\mu}$ ($c_{n,\mu}$) is a creation (an annihilation) operator of a particle with index $\mu$ (we assume a unit cell is composed of $M$ degrees of freedom, hence, $\mu=1,\ldots,M$) in the $n$-th unit cell.  

We denote $H(\beta) = h+V\beta+W \beta^{-1}$ as a matrix-valued rational function of $\beta$. Consider the generalized eigenproblem $[\omega - H(\beta)]x =0$. It can be seen that
$\det[\omega - H(\beta)]= 0$ has $2M$ $\beta$ solutions $\beta_i,i=1,\ldots,2M$ and we denote $x_i$ be the corresponding nullvector of $\beta_i$. In terms of $\beta_i,x_i$, we obtain the following formula of $G_{11}(\omega)$ in the thermodynamics limit (see Appendix \ref{sec: Green_app} for detail),
\begin{align}
    & \lim_{N \rightarrow \infty} G_{11}(\omega) \notag \\
    = & (x_1,\ldots,x_M)(\omega) \text{diag}(\beta_1^{-1},\ldots,\beta_M^{-1}) (x_1,\ldots,x_M)^{-1}(\omega) ,
    \label{eq: main_G11_therm}
\end{align}
where $\beta_1,\ldots,\beta_M$ are $M$ suitable chosen $\beta$ solutions and $|\beta_1| \geqslant |\beta_2| \geqslant \ldots \geqslant |\beta_{M}| > |\beta_{M+1}| \geqslant \ldots \geqslant |\beta_{2M}|$ for general non-Hermitian systems without any discrete symmetry or constraint.

As the first order perturbation, $x_i(\omega)=x_i(0)+\omega \delta x_i$ depends linearly on $\omega$. Hence, if $\det (x_1(0),\ldots,x_M(0)) =0$, $(x_1,\ldots,x_M)^{-1}(\omega)$ term in Eq.~(\ref{eq: main_G11_therm}) contributes a nonzero coefficient matrix of the $\frac{1}{\omega}$ term of $\lim_{N \rightarrow \infty} G_{11}(\omega)$. Otherwise, the coefficient matrix of the $\frac{1}{\omega}$ term of $\lim_{N \rightarrow \infty} G_{11}(\omega)$ is the zero matrix, and $\lim_{N \rightarrow \infty} G_{11}(\omega)$ has no poles at $\omega=0$.

As the first example, we consider the simplest model with the sublattice symmetry, the number of the internal degrees of freedom $M=2$ in this model. The Hamiltonian
\begin{align}
    H(\beta) = \begin{pmatrix}
        0 & H_-(\beta) \\
        H_+(\beta) & 0
    \end{pmatrix}
    \label{eq: H_SLS}
\end{align},
where $H_+(\beta)$ and $H_-(\beta)$ are $1 \times 1$ matrices (scalar). Without loss of generality, $H_{\pm}(\beta)$ can be factored as $H_{\pm}(\beta)=a_{\pm} \frac{(\beta-\beta_{\pm,1})(\beta-\beta_{\pm,2})}{\beta}$ with $\beta_{\pm,1}$ and $\beta_{\pm,2}$ being zeros of $H_{\pm}(\beta)$. $\beta_i(0)=\beta_{\pm,1},\beta_{\pm,2}$ are solutions of $\det\left(H(\beta)\right)=0$, and the corresponding nullvectors $x_i(0)$ of $H(\beta_i(0))$ are 
\begin{align}
    x_+ = (1,0)^T \text{ for } \beta(0)=\beta_{+,1} \text{ and } \beta_{+,2}, \notag \\
    x_- = (0,1)^T \text{ for } \beta(0)=\beta_{-,1} \text{ and } \beta_{-,2}.
\end{align}
By above discussion, if $\beta_{+,1},\beta_{+,2}$ are two $\beta$ solutions with the largest magnitudes, i.e., $|\beta_{+,1}|,|\beta_{+,2}|>|\beta_{-,1}|,|\beta_{-,2}|$, $(x_1(0),x_2(0)) = (x_+,x_+)$ is singular, and the coefficient matrix of the the $\frac{1}{\omega}$ term of $\lim_{N \rightarrow \infty} G_{11}(\omega)$ term has the form of $\begin{pmatrix}
    0 & 0 \\
    0 & A_{22}
\end{pmatrix}$. For this case, by using Eq.~(\ref{eq: SLS_BBC}) or Eq.~(\ref{eq: Topo_A}), the topological invariant $\lim_{\omega \rightarrow 0} -\frac{\Tr (\Gamma r(\omega))}{2}=-1$. If $|\beta_{-,1}|,|\beta_{-,2}|>|\beta_{+,1}|,|\beta_{+,2}|$, $(x_1(0),x_2(0)) = (x_-,x_-)$ is singular, and the coefficient matrix of the the $\frac{1}{\omega}$ term of $\lim_{N \rightarrow \infty} G_{11}(\omega)$ term has the form of $\begin{pmatrix}
    A_{11} & 0 \\
    0 & 0
\end{pmatrix}$. For this case, the topological invariant $\lim_{\omega \rightarrow 0} -\frac{\Tr (\Gamma r(\omega))}{2}=1$. For other cases, $(x_1(0),x_2(0))=(x_+,x_-)$ is non-singular and the topological invariant $\lim_{\omega \rightarrow 0} -\frac{\Tr (\Gamma r(\omega))}{2}=0$ which implies the system is in the trivial phase. From the above discussion, it can be seen that $\text{rank} \left(x_1(0),x_2(0)\right)$ changes when $x_2(0)$ and $x_3(0)$ are exchanged, which implies that the order of magnitudes of $\beta_2(0)$ and $\beta_3(0)$ is reversed. Hence, the topological phase transition occurs at $|\beta_2(0)| = |\beta_3(0)|$, which is understood as the bulk energy gap closing point from the bulk energy spectrum viewpoint by the GBZ approach.

\subsection{Critical topological phase}
\label{sec: critical_topo}
We consider a four-band model with the sublattice symmetry as the second example, which illustrates a new unstable topological phase, dubbed the critical topological phase. The Hamiltonian is the same form as in Eq.~(\ref{eq: H_SLS}), where 
\begin{align}
    & H_{\pm}(\beta) \notag \\
    = & \begin{pmatrix}
        a_{\pm} \frac{(\beta-\beta_{\pm,a,1})(\beta-\beta_{\pm,a,2})}{\beta} & c \\
        c & b_{\pm} \frac{(\beta-\beta_{\pm,b,1})(\beta-\beta_{\pm,b,2})}{\beta}
    \end{pmatrix}.
\end{align}
In our model, $|\beta_{+,a,1}|,|\beta_{+,a,2}|>|\beta_{-,a,1}|,|\beta_{-,a,2}|>|\beta_{+,b,1}|,|\beta_{+,b,2}|>|\beta_{-,b,1}|,|\beta_{-,b,2}|$. In the followings, we show that the Green's function of this Hamiltonian has huge difference between $c=0$ and $c\neq 0$ cases. In coupled-resonators systems \cite{peng2014loss, Longhi2015, doi:10.1126/science.aay1064}, the off-diagonal coupling $c$ of $H_{\pm}(\beta)$ represents the coupling strength between two resonators of the same unit cell .

When $c=0$, the Hamiltonian $H$ has a constraint that it is the direct sum of two independent two-band Hamiltonians with the sublattice symmetry $H=H_a \oplus H_b$, where the index $a$ represents the first diagonal element of $H_{\pm}$ and the index $b$ represents the second diagonal element of $H_{\pm}$. Hence, $\beta_1,\ldots,\beta_M$ should be chosen from $\beta_{\pm,a,i}$ and $\beta_{\pm,b,i}$ separately, i.e., it should be chosen by selecting two $\beta$ with largest magnitude from $\beta_{\pm,a,i}$ and selecting two $\beta$ with largest magnitude from $\beta_{\pm,b,i}$. Hence, $x_1(0),\ldots,x_M(0)$ in Eq.~(\ref{eq: main_G11_therm}) should be nullvectors corresponding to $\beta_{+,a,1},\beta_{+,a,2},\beta_{+,b,1},\beta_{+,b,2}$. Since $(x_1(0),\ldots,x_M(0))$ is singular, this system is the topological phase, and it can be calculated that the topological invariant $\lim_{\omega \rightarrow 0} -\frac{\Tr (\Gamma r(\omega))}{2}=2$.

When $c \neq 0$, the $\beta$ solutions of $\det[H(\beta)]=0$ deviate slightly from $\beta_{\pm,a,i}$ and $\beta_{\pm,b,i}$ while the order of their magnitudes keeps invariant, therefore, we denote them by $\beta'_{\pm,a,i}$ and $\beta'_{\pm,b,i}$ with the same subscript indices. The Hamiltonian $H$ has no constraint, and $\beta_1,\ldots,\beta_M$ should be chosen as $M$ largest $\beta$ from the union of $\beta'_{\pm,a,i}$ and $\beta'_{\pm,b,i}$. $(x_1(0),\ldots,x_M(0))$ should be nullvectors corresponding to $\beta'_{+,a,1}(\omega),\beta'_{+,a,2}(\omega),\beta'_{-,a,1}(\omega),\beta'_{-,a,2}(\omega)$ which is non-singular. Hence, this system is the trivial phase with $\lim_{\omega \rightarrow 0} -\frac{\Tr (\Gamma r(\omega))}{2}=0$.

\begin{figure}
    \includegraphics[width=0.8\columnwidth]{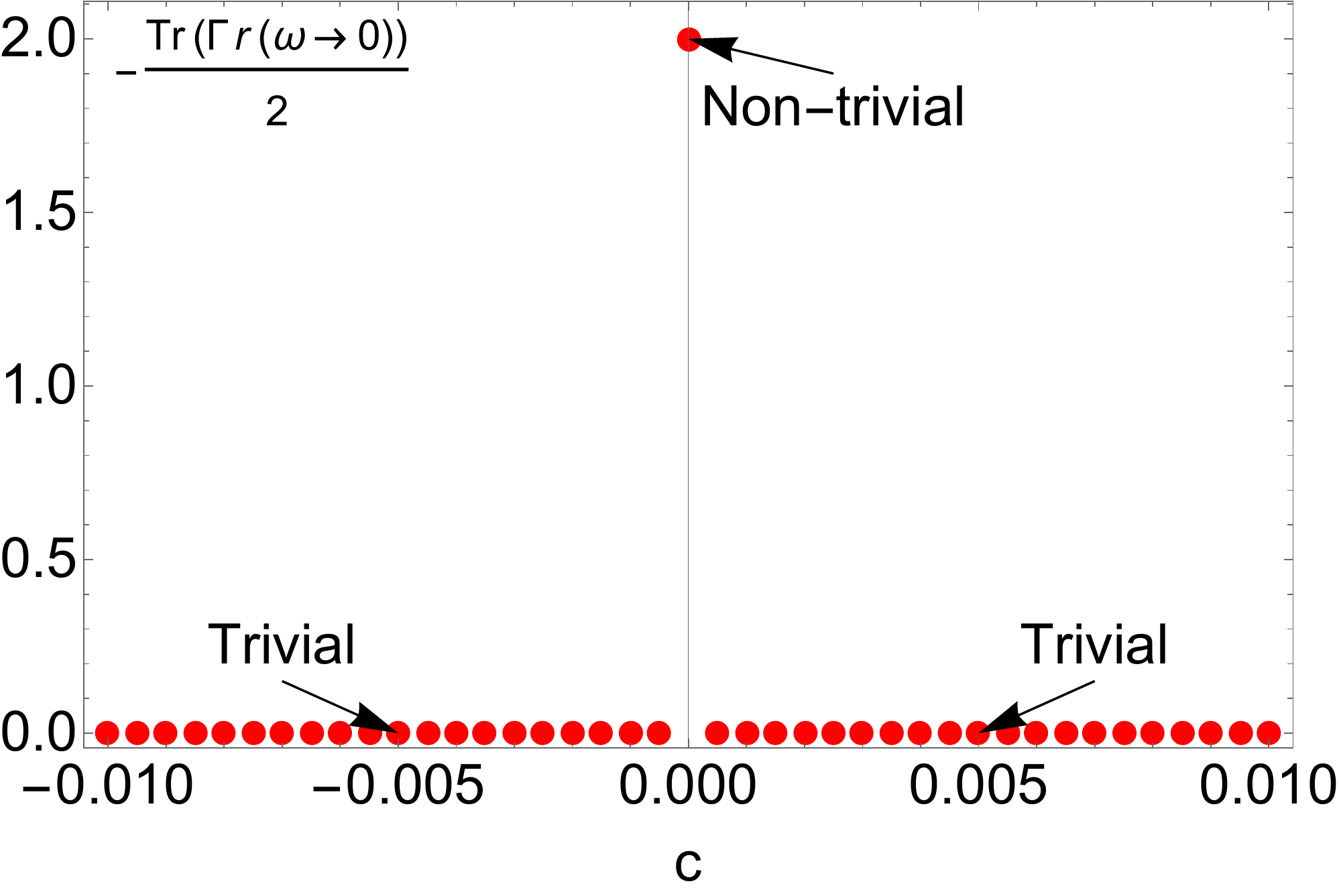}
    \caption[]{Numerical result of the topological invariant $\lim_{\omega \rightarrow 0} -\frac{\Tr (\Gamma r(\omega))}{2}$ varies with the off-diagonal coupling $c$ of $H_{\pm}(\beta)$. It shows the critical topological phase transition. Here, $a_{\pm}=b_{\pm}=1$, $\beta_{+,a,1}=20,\beta_{+,a,2}=10,\beta_{-,a,1}=6,\beta_{-,a,2}=5,\beta_{+,b,1}=2,\beta_{+,b,2}=1,\beta_{-,b,1}=\frac{1}{5},\beta_{-,b,2}=\frac{1}{10}$. At $c=0$, there exists localized zero energy modes at two boundaries, while there is no localized zero energy modes at two boundaries when $c \neq 0$.}
    \label{fig: 4band-topo-trans}
\end{figure}

The different selection results of $(x_1,\ldots,x_M)$ and $(\beta_1,\ldots,\beta_M)$ for $c=0$ and $c\neq 0$ cases imply a dramatic change of the Green's function even with only a small change in $c$. Furthermore, The discontinuity of $(x_1,\ldots,x_M)$ and $(\beta_1,\ldots,\beta_M)$ selection at $c=0$ can cause a new type of topological phase transition of non-Hermitian systems as shown in Fig.~\ref{fig: 4band-topo-trans}. The system is in the topological phase only at single parameter point $c=0$ (there exists localized zero energy modes at two boundaries), and this topological phase (it is called topological phase since the existence of gapless localized boundary states) is not robust since it can be broken by a very small variation of $c$ although the bulk energy gap of the system at $c=0$ is large compared to the variation of $c$. In addition, between two gapped phases ($c=0$ and $c\neq 0$), there is no gapless phase at the topological transition point.
The discontinuity of $(x_1,\ldots,x_M)$ and $(\beta_1,\ldots,\beta_M)$ selection at $c=0$ does not exist in Hermitian cases, which is detailed in Appendix \ref{sec: Herm_c}. Therefore, this type of topological phase transition only occurs in non-Hermitian systems. We also want to mention that from the viewpoint of the GBZ approach, this type of topological phase transition is caused by a dramatic change of the GBZ, which is dubbed as the critical skin effect \cite{PhysRevLett.125.226402}. Here, we call this new type of topological phase transition the critical topological phase transition and the topological phase at $c=0$ the critical topological phase.

\section{$\mathbb{Z}_2$ topological phase} \label{sec: TR}
In this section, we develop our method in a non-Hermitian $\mathbb{Z}_2$ topological phase. A well-known $\mathbb{Z}_2$ topological phase is protected by the time-reversal symmetry (TRS) \cite{science.1133734, PhysRevLett.95.226801, PhysRevLett.95.146802, PhysRevB.74.195312, science.1148047, PhysRevB.76.045302, PhysRevLett.98.106803, PhysRevB.78.195424, RevModPhys.88.035005, PhysRevX.9.041015}. One of the important features of the time-reversal invariant system is the Kramer degeneracy. For non-Hermitian Hamiltonians, the symplectic class with TRS$^\dagger$ defined by \cite{PhysRevX.9.041015, PhysRevB.101.195147}
\begin{align}
    \mathcal{U}_T H^T \mathcal{U}_T^{-1} = H, \; \; \mathcal{U}_T \mathcal{U}_T^{*} =-1,
\end{align}
with a unitary matrix $\mathcal{U}_T$ representing an analogy of the TRS has the Kramer degeneracy. Non-Hermitian systems with TRS$^\dagger$ is proved to have paired subGBZs \cite{PhysRevB.101.195147}. For multi-subGBZs systems, the GBZ-based approach to calculate the topological invariant is cumbersome, hence, we would like to introduce the scattering matrix method in non-Hermitian systems with TRS$^\dagger$.

For 1D systems with only TRS$^\dagger$, there is no topological phase. Here, we consider 1D non-Hermitian systems with both the sublattice symmetry and TRS$^\dagger$, i.e., class D with the sublattice symmetry (class $(S_{-},D)$ in TABLE VII. of Ref.~\cite{PhysRevX.9.041015}). The Hamiltonian $H$ satisfies 
\begin{align}
    \Gamma H \Gamma^{-1} = -H, \; \; 
    \mathcal{U}_T H^T \mathcal{U}_T^{-1} = H.
\end{align}
The system also has a particle-hole symmetry $\mathcal{U}_C = \mathcal{U}_T \Gamma$. The Green's function obeys 
\begin{align}
    \Gamma G(\omega) \Gamma^{-1} & = -G(-\omega), \notag \\
    \mathcal{U}_T G(\omega) \mathcal{U}_T^{-1} & = G^{T}(\omega), \notag \\
    \mathcal{U}_C G(\omega) \mathcal{U}_C^{-1} & = -G^{T}(-\omega).
    \label{eq: TC_G}
\end{align}
By Eq.~(\ref{eq: r_G}), when $\omega \rightarrow 0$,
\begin{align}
    \Gamma r \Gamma^{-1} & = r^{-1}, \notag \\
    \mathcal{U}_T r \mathcal{U}_T^{-1} & = r^T , \notag \\
    \mathcal{U}_C r \mathcal{U}_C^{-1} & = (r^{-1})^T.
\end{align}
Particle-hole symmetry operator $\mathcal{U}_C$ is a unitary operator obeys $\mathcal{U}_C \mathcal{U}_C^{*} = 1$ (provided $\{\mathcal{U}_T, \Gamma \}=0$), which impiles $\mathcal{U}_C = \mathcal{U}_C^T$. Therefore, there exists an operator $\mathcal{V}_C$ such that $\mathcal{U}_C = \mathcal{V}_C \mathcal{V}_C^T$. When $\omega \rightarrow 0$, it follows that 
\begin{align}
    \left[ \mathcal{V}_C^{\dagger} \Gamma r \mathcal{V}_C \right]^T  = -\mathcal{V}_C^{\dagger} \Gamma r \mathcal{V}_C.
\end{align}
Since $\mathcal{V}_C^{\dagger} \Gamma r \mathcal{V}_C$ is antisymmetric, the Pfaffian of $\mathcal{V}_C^{\dagger} \Gamma r \mathcal{V}_C$ is well-defined and we simply denote it by $\text{Pf}\left( \mathcal{V}_C^{\dagger} \Gamma r \mathcal{V}_C \right)$.

Now, we show that $\text{Pf}\left( \mathcal{V}_C^{\dagger} \Gamma r \mathcal{V}_C \right)$ is quantized. By the sublattice symmetry, $(\Gamma \, r(0))^2 = \mathbb{I}$, it follows that $\det(\mathcal{V}_C^{\dagger} \Gamma r \mathcal{V}_C) = \det(\Gamma r) = \pm 1$. Furthermore, by TRS$^\dagger$, $\det(\Gamma r) = 1$. Hence, when $\omega \rightarrow 0$,
\begin{align}
    \text{Pf}\left( \mathcal{V}_C^{\dagger} \Gamma r \mathcal{V}_C \right)^2 & = \det(\mathcal{V}_C^{\dagger} \Gamma r \mathcal{V}_C) = 1 \notag \\
    \Rightarrow \text{Pf}\left( \mathcal{V}_C^{\dagger} \Gamma r \mathcal{V}_C \right) & = \pm 1.
\end{align}
Therefore, $\text{Pf}\left( \mathcal{V}_C^{\dagger} \Gamma r \mathcal{V}_C \right)$ is a well-defined topological invariant.

\subsection{Establish the BBC}
\label{sec: BBC_TRS}
For the same reason as we discuss the BBC of systems protected by the sublattice symmetry, after taking the thermodynamics limit, the diagonal block of the real space Green's function at the first unit cell can be written as 
\begin{align}
    G_{11}(\omega) = \frac{A}{\omega} + (\text{regular terms at }\omega=0),
\end{align}
where $G_{11}(\omega)$ and $A=\left( \sum_{n : E_n =0}\braket{1 | n R}\braket{n L| 1} \right)$ are $M \times M$ matrices when the number of the internal degree of freedom is $M$ at each unit cell of the system. 

By Eq.~(\ref{eq: Topo_A_TRS}) in Appendix \ref{sec: Topo_inv_G}, we obtain the BBC in the thermodynamics limit 
\begin{align}
    (-1)^{\frac{M}{2}\left( \frac{M}{2} -1 \right)/2} \text{Pf}(i \mathcal{V}_C^{\dagger} \Gamma r(0) \mathcal{V}_C) = (-1)^p,
    \label{eq: TRS_BBC}
\end{align}
where $p$ is the number of Kramer pairs of non-zero eigenvalues of $A=\sum_{n : E_n =0}\braket{1 | n R}\braket{n L| 1}$. This BBC is understood as followings, the left hand side of Eq.~(\ref{eq: SLS_BBC}) represents the bulk topological invariant and the right hand side of Eq.~(\ref{eq: TRS_BBC}) encodes the information about the number of zero energy modes ``localized'' at the left end.

\begin{figure} 
    \centering
    \subfigure[]{\includegraphics[width=0.95\columnwidth]{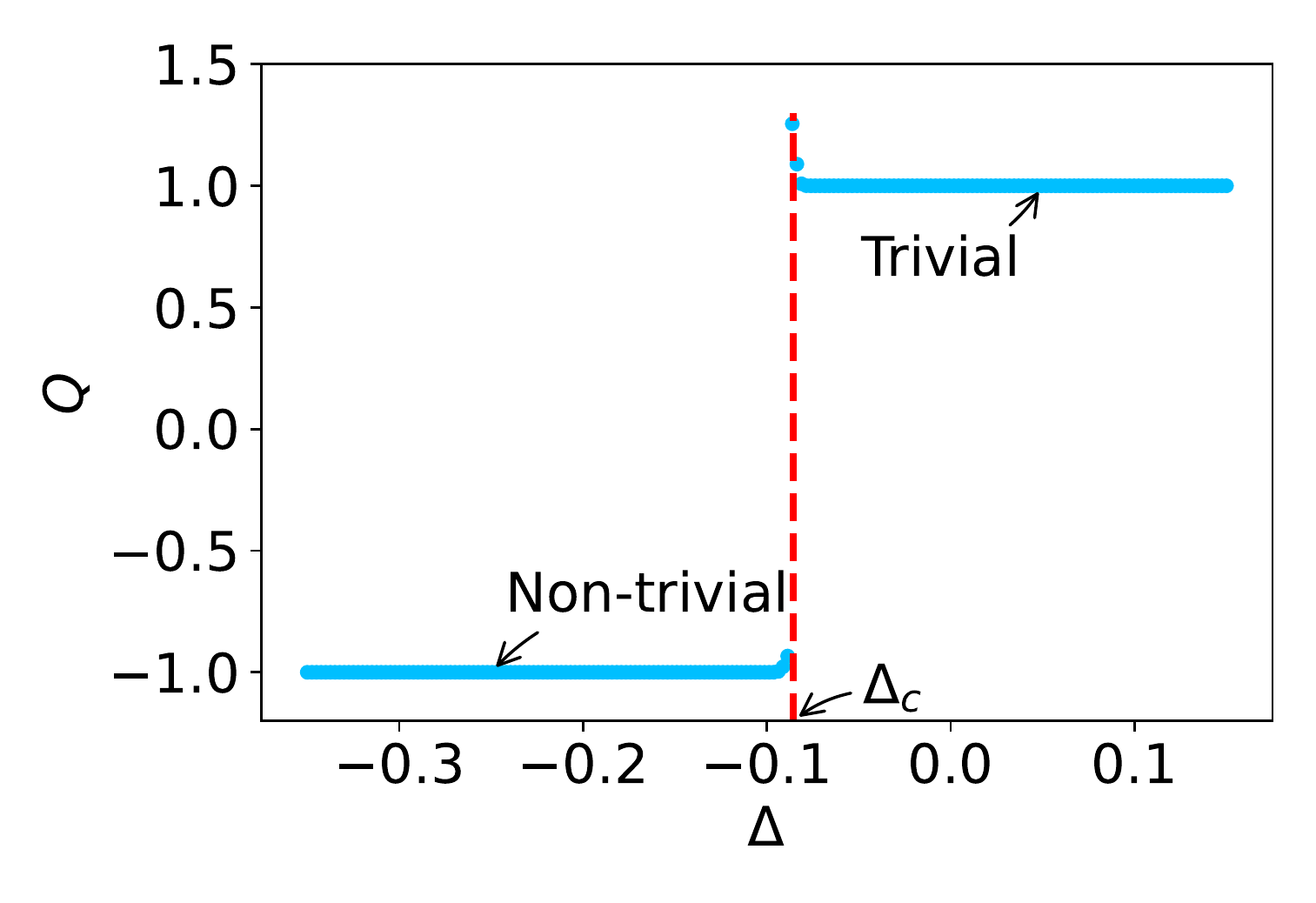}}
    \subfigure[]{\includegraphics[width=0.45\columnwidth]{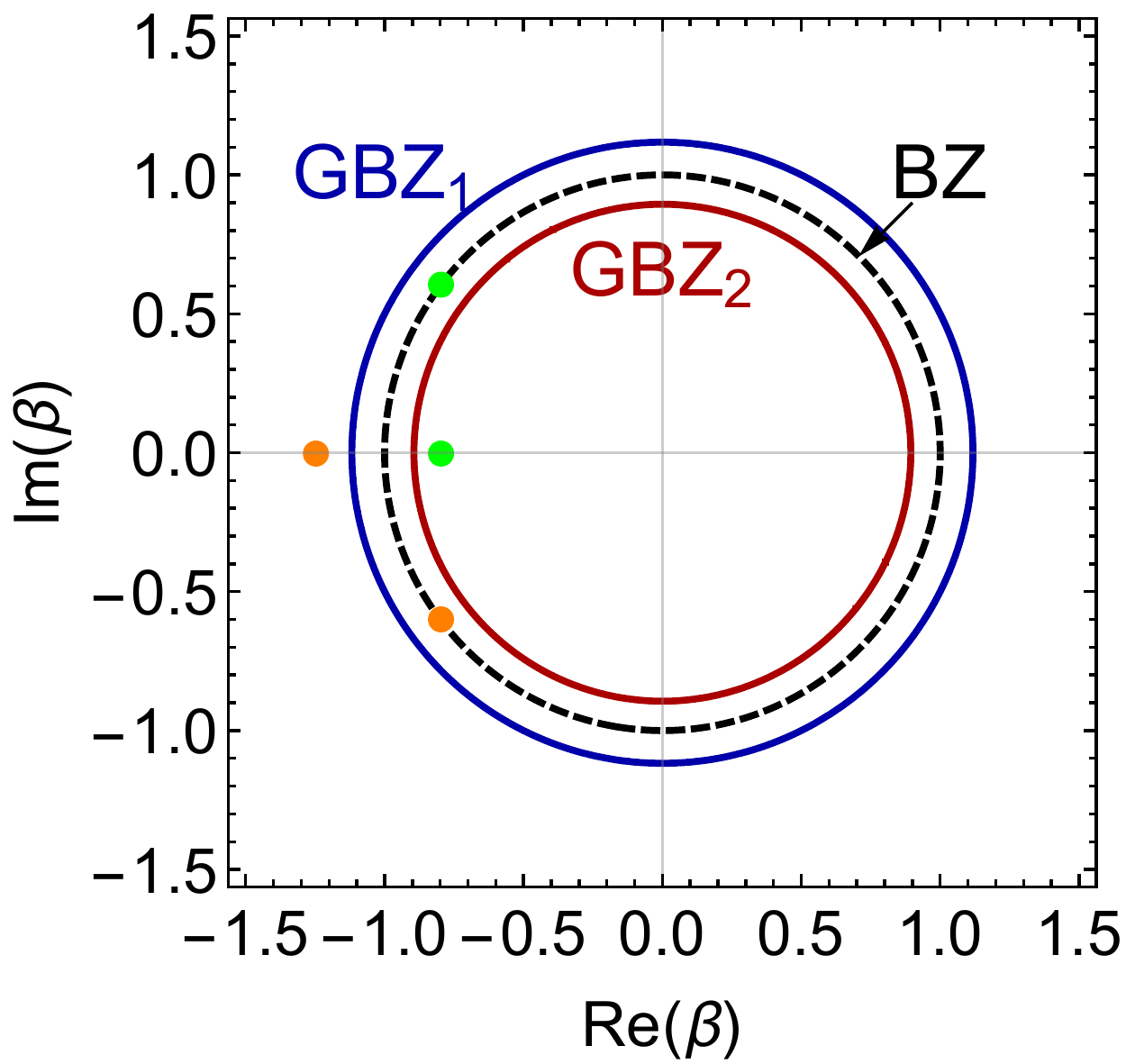}}
    \subfigure[]{\includegraphics[width=0.45\columnwidth]{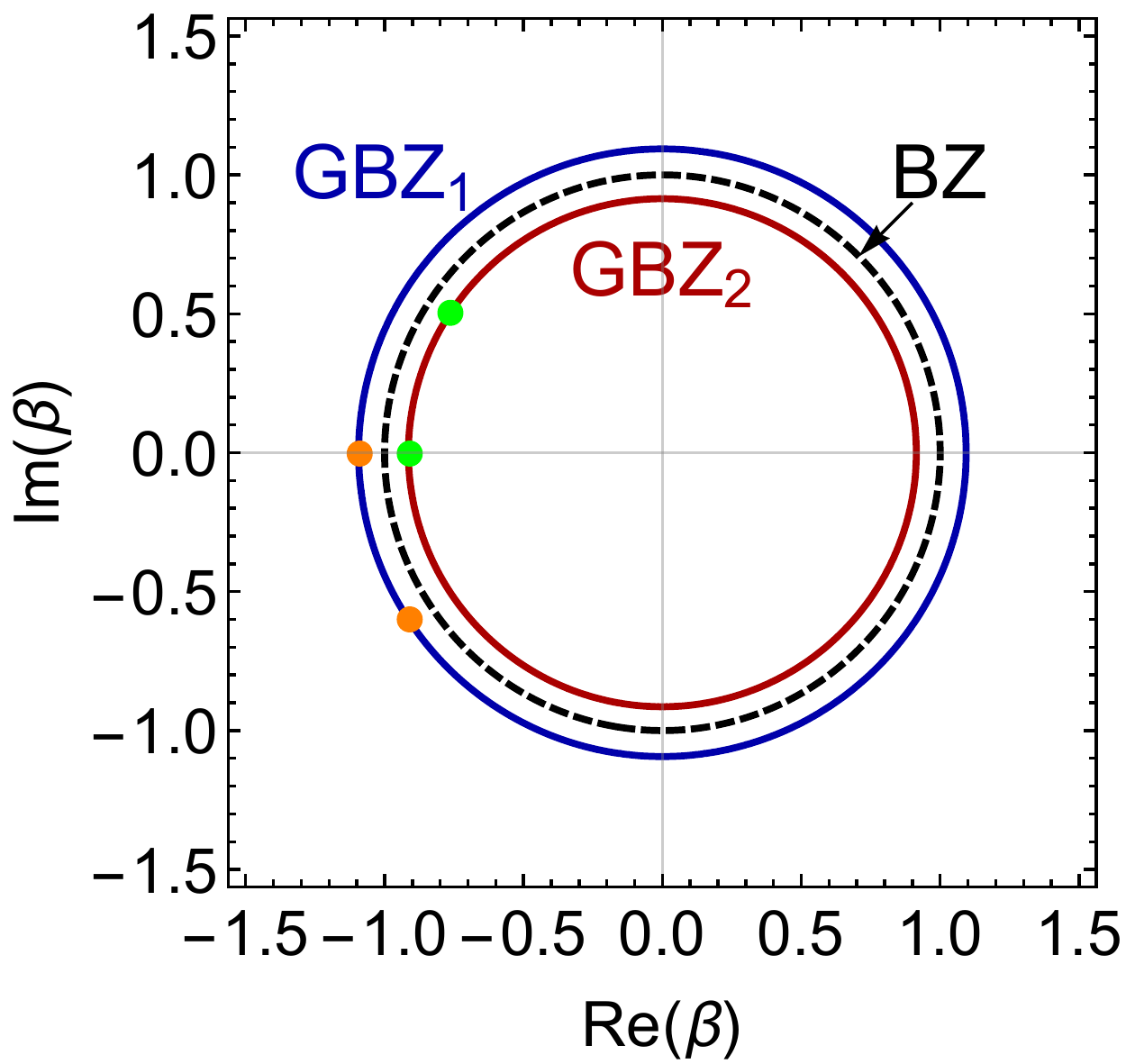}}
    \caption[TRS_transition]{Here, $t=u=1$ and $\gamma=6/5$ in Eq.~(\ref{eq: TR_model}). (a) The topological invariant $Q = (-1)^{\frac{M}{2}\left( \frac{M}{2} -1 \right)/2} \text{Pf}(i \mathcal{V}_C^{\dagger} \Gamma r(0) \mathcal{V}_C)$ as a function of the parameter $\Delta$. (b) $\beta$ spectrum when $\Delta = -0.2<\Delta_c$. $\beta$ spectrum includes bulk states $\beta$ spectrum (two subGBZs) and possible boundary states $\beta$ spectrum (discrete zeros of $\det[H(\beta)]$). When $\Delta>\Delta_c$, the $\beta$ spectrum is similar, but any linear combination of states with possible boundary states $\beta$ spectrum can not satisfy the boundary condition, which impiles that there is no boundary state.
    (c) $\beta$ spectrum when $\Delta = \Delta_c$. Boundary states $\beta$ spectrum are on two subGBZs, which shows that boundary states are merged into bulk and $\Delta_c$ is the topological transition point under the OBC.}
    \label{fig: TRS_transition}
\end{figure}

As an example, we consider the following non-Hermitian Bloch Hamiltonian 
\begin{align}
    H(k) = \begin{pmatrix}
        0 & D_1(k) \\
        D_2(k) & 0 
    \end{pmatrix},
    \label{eq: TR_model}
\end{align}
where $D_1(k) = t \sin k \, \sigma_x + \left( \Delta + u + u \cos k + i \frac{\gamma}{2} \right) \sigma_y$ and $D_2(k) = t \sin k \, \sigma_x + \left( \Delta + u + u \cos k \right) \sigma_y$. The sublattice symmetry operator $\Gamma = \sigma_z \otimes \mathbb{I}_{2 \times 2}$ and the TRS$^\dagger$ operator $\mathcal{U}_T = -i \sigma_y \otimes \mathbb{I}_{2 \times 2}$. Under the periodic boundary condition (PBC), the bulk energy gap closes at $\Delta=0$ and $\Delta=\pm \sqrt{t^2-(\frac{\gamma}{2})^2}-t$. However, under the open boundary condition (OBC), the topological transition points change to $\Delta = \Delta_c$  and $\Delta= -2 -\Delta_c$ due to the occurrence of the NHSE ($\Delta_c= \frac{1}{10} \left(\sqrt{2 \left(\sqrt{2581}-9\right)}-10\right)$ for $t=u=1$ and $\gamma=6/5$). We show the topological invariant $Q=\text{Pf}\left( \mathcal{V}_C^{\dagger} \Gamma r \mathcal{V}_C \right)$ and $\beta$ spectrum near $\Delta = \Delta_c$ in Fig.~\ref{fig: TRS_transition}. Here, $\beta$ spectrum includes bulk states $\beta$ spectrum (two subGBZs) and possible boundary states $\beta$ spectrum (discrete zeros of $\det[H(\beta)]$). For gapless phases at the topological transition point, discrete boundary states $\beta$ spectrum are on GBZ since at this point boundary states are merged into bulk. The topological non-trivial phase and the topological trivial phase can usually be distinguished by the existence of the boundary states. For the topological non-trivial phase, the linear combination of states with boundary states $\beta$ spectrum can satisfy the boundary condition while the boundary condition can not be satisfied in the topological trivial phase. 

\subsection{Analytical computation of the topological invariant}
\label{sec: TRS_analy}
\begin{figure} 
    \centering
    \includegraphics[width=0.95\columnwidth]{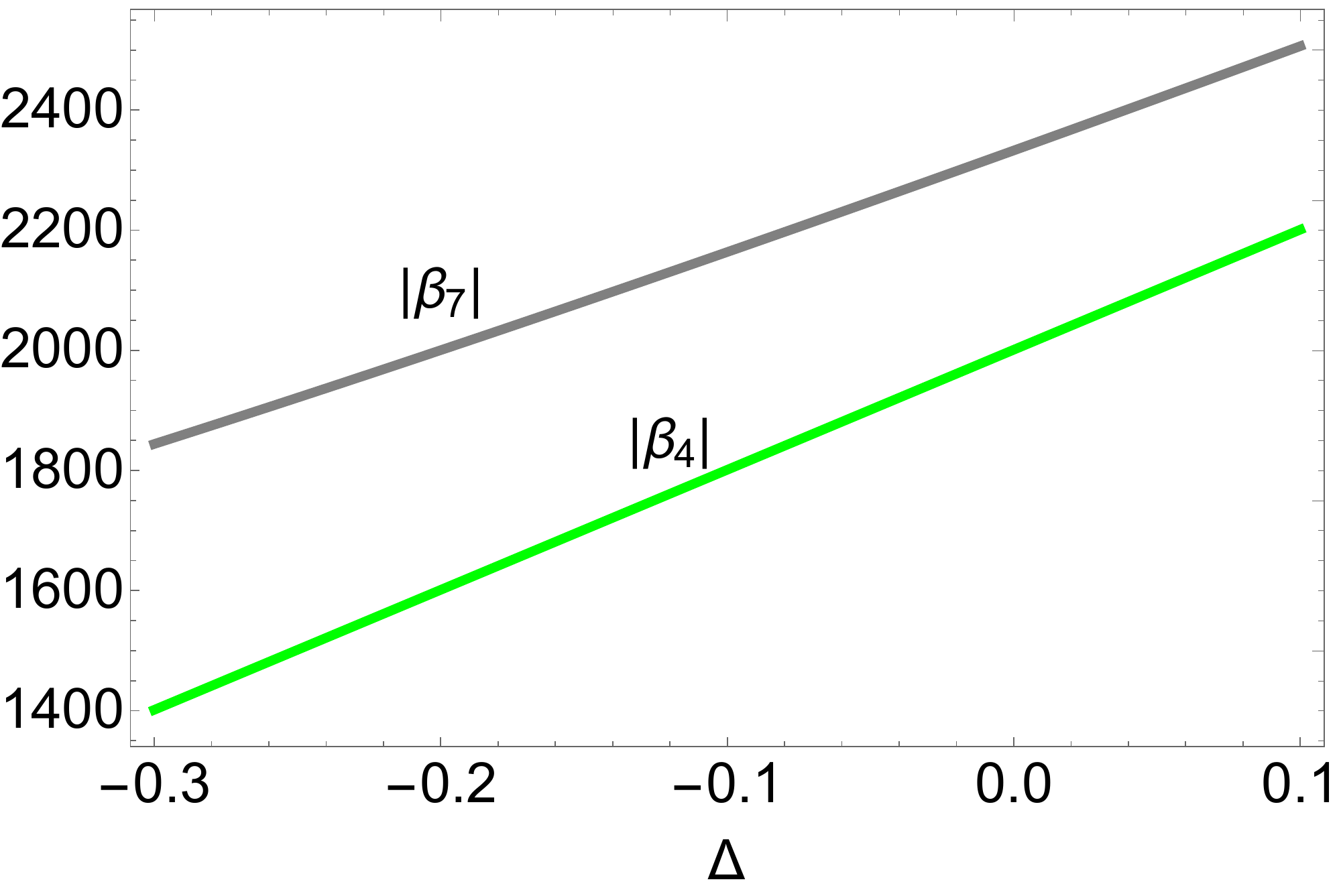}
    \includegraphics[width=0.95\columnwidth]{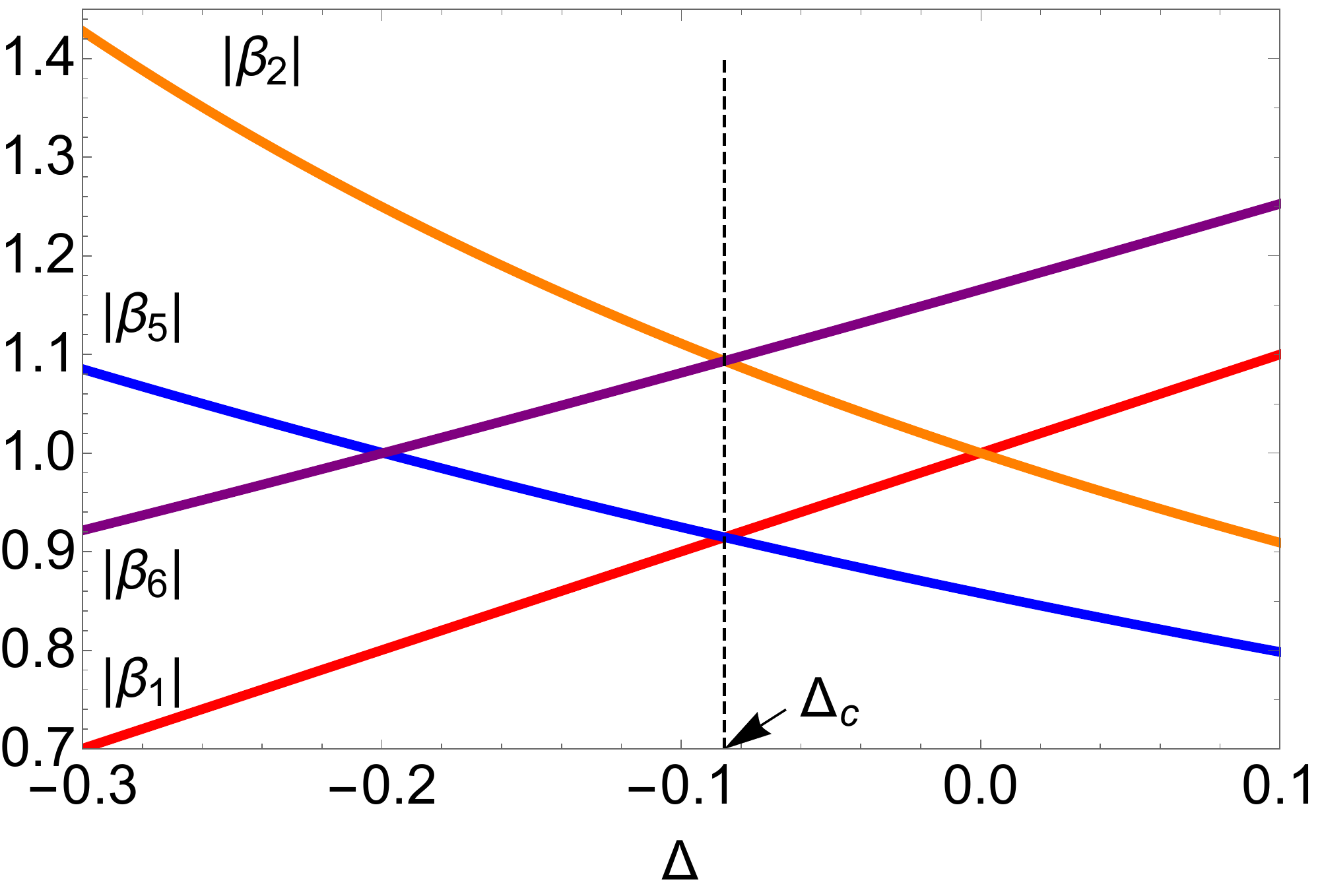}
    \includegraphics[width=0.95\columnwidth]{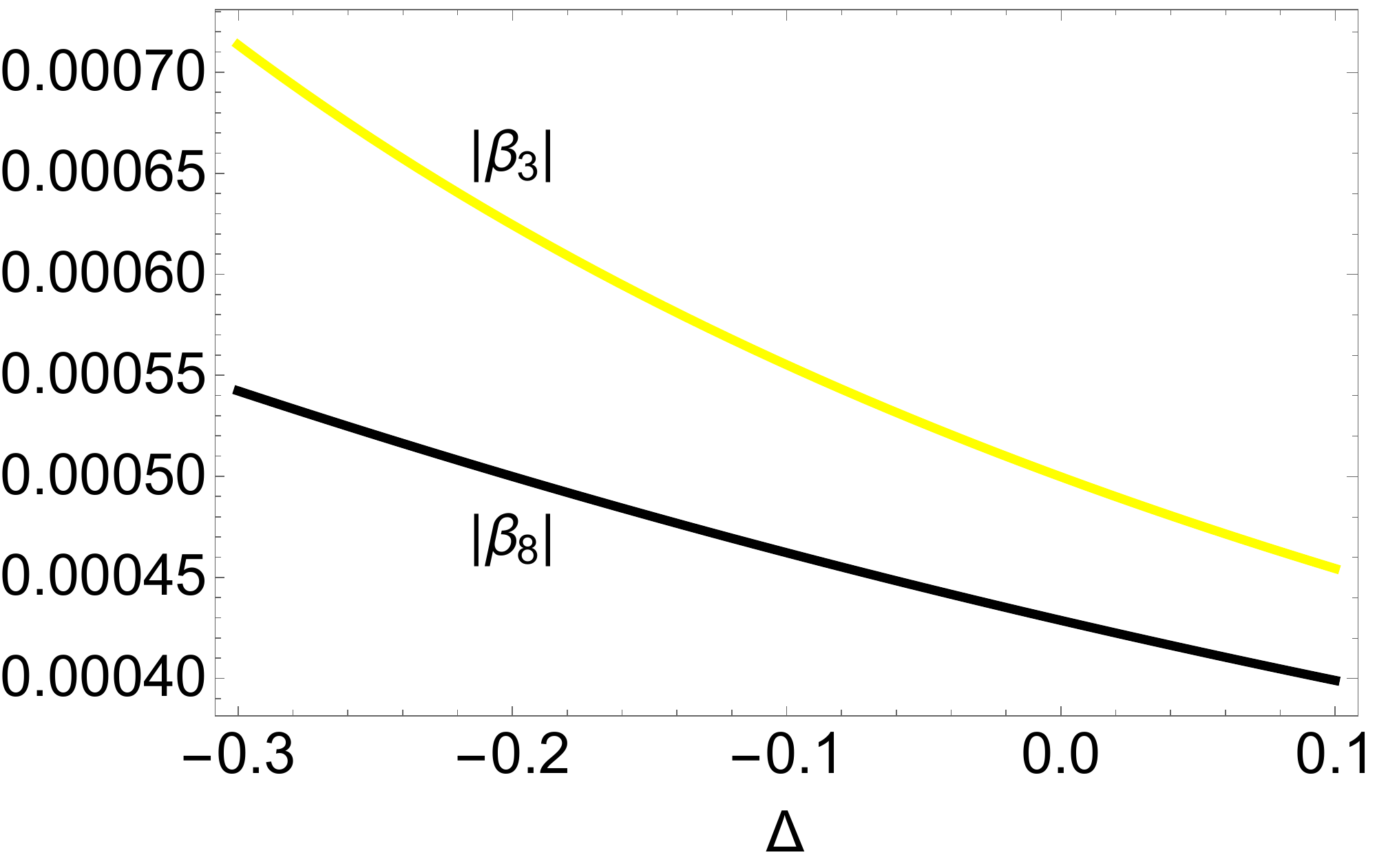}
    \caption[beta_Abs]{The magnitudes of $\beta$ solutions labeled in Eq~(\ref{eq: beta_y}). Here, $t=u=1$ and $\gamma=6/5$. Crossings at $\Delta=\Delta_c$ shows that $\Delta=\Delta_c$ is the topological transition point.}
    \label{fig: beta_Abs}
\end{figure}

In this section, we apply the method developed in Sec.~\ref{sec: analy} to the model (\ref{eq: TR_model}) with TRS$^\dagger$. In this model, hopping amplitudes in Eq.~(\ref{eq: main_Ham}) are
\begin{align}
    h = \begin{pmatrix}
        0 & (\Delta+u+i\frac{\gamma}{2})\sigma_y \\
        (\Delta+u) \sigma_y & 0
    \end{pmatrix}, \notag \\
    V = \begin{pmatrix}
        0 & -\frac{t}{2i}\sigma_x+\frac{u}{2}\sigma_y \\
        -\frac{t}{2i}\sigma_x+\frac{u}{2}\sigma_y & 0
    \end{pmatrix}, \notag \\
    W = \begin{pmatrix}
        0 & \frac{t}{2i}\sigma_x+\frac{u}{2}\sigma_y \\
        \frac{t}{2i}\sigma_x+\frac{u}{2}\sigma_y & 0
    \end{pmatrix}.
\end{align}
Consider the generalized eigenequation $H(\beta_i) y_i = 0 , \; i=1,\ldots,2M$ with $H(\beta_i)=h + V \beta_i + W \beta_i^{-1}$. The solution is 
\begin{widetext}
\begin{align}
    \beta_1 & = -\frac{\Delta +\sqrt{\Delta ^2+t^2+2 \Delta  u}+u}{t+u} , \; y_1 = (1,0,0,0)^T , \notag \\
    \beta_2 & = \frac{\Delta -\sqrt{\Delta ^2+t^2+2 \Delta  u}+u}{t-u} , \; y_2 = (0, 1, 0, 0)^T , \notag \\
    \beta_3 & = \frac{\Delta -\sqrt{\Delta ^2+t^2+2 \Delta  u}+u}{t+u}, \; y_3 = (1, 0, 0, 0)^T , \notag \\
    \beta_4 & = \frac{\Delta +\sqrt{\Delta ^2+t^2+2 \Delta  u}+u}{t-u}, \; y_4 = (0, 1, 0, 0)^T , \notag \\
    \beta_5 & = \frac{i \gamma +2 \Delta -\sqrt{4 t^2+(2 \Delta +i \gamma ) (i \gamma +2 \Delta +4 u)}+2 u}{2 (t-u)}, \;  y_5 = (0, 0, 0, 1)^T , \notag \\
    \beta_6 & = -\frac{i \gamma +2 \Delta +\sqrt{4 (t-u) (t+u)+(i \gamma +2 \Delta +2 u)^2}+2 u}{2 (t+u)} , \; y_6 = (0, 0, 1, 0)^T , \notag \\
    \beta_7 & = \frac{i \gamma +2 \Delta +\sqrt{4 t^2+(2 \Delta +i \gamma ) (i \gamma +2 \Delta +4 u)}+2 u}{2 (t-u)}, \; y_7 = (0, 0, 0, 1)^T , \notag \\
    \beta_8 & = \frac{-i \gamma -2 \Delta +\sqrt{4 (t-u) (t+u)+(i \gamma +2 \Delta +2 u)^2}-2 u}{2 (t+u)} , \; y_8 = (0, 0, 1, 0)^T.
    \label{eq: beta_y}
\end{align}
\end{widetext}
Notice that due to the existence of TRS$^{\dagger}$, $\det[H(\beta)]$ can be factored as $\det[H(\beta)] = \frac{f(\beta) g(\beta)}{\beta^4}$ with $f(\beta)$ and $g(\beta)$ being polynomials. $\beta_{2i-1},i=1,\ldots,4$ are solutions of $f(\beta)=0$ and $\beta_{2i},i=1,\ldots,4$ are solutions of $g(\beta)=0$ respectively, and $\beta_{2i-1}=\frac{1}{\beta_{2i}}$.
Similar to the second example discussed in Sec.~\ref{sec: critical_topo} with $c=0$, contributions of terms involving $\beta_{2i-1},i=1,\ldots,4$ and $\beta_{2i},i=1,\ldots,4$ are independent. Thus dominant terms of Eq.~(\ref{eq: G11_by_parts}) are contributed by terms involving $\beta_{2i-1},i=1,\ldots,4$ and $\beta_{2i},i=1,\ldots,4$ independently, and $\beta_{2i-1},i=1,\ldots,4$ and $\beta_{2i},i=1,\ldots,4$ should be ordered separately to determine $\beta_1(0),\ldots,\beta_M(0)$ and $x_1(0),\ldots,x_M(0)$ in Eq.~(\ref{eq: main_G11_therm}). As shown in Fig.~\ref{fig: beta_Abs}, we plot the magnitudes of $\beta_i$ vary with the parameter $\Delta$ near $\Delta=\Delta_c$ for $t=u=1$ and $\gamma=6/5$. By Fig.~\ref{fig: beta_Abs}, for $\Delta < \Delta_c$, $(x_1(0),\ldots,x_M(0))$ in Eq.~(\ref{eq: main_G11_therm}) is $(y_7,y_5,y_4,y_2)$ which is singular, and the system is in the topological phase. For $\Delta > \Delta_c$, $(x_1(0),\ldots,x_M(0))$ in Eq.~(\ref{eq: main_G11_therm}) is $(y_7,y_1,y_4,y_6)$ which is non-singular, and the system is in the topological trivial phase.  

Note that the topological phase transition occurs at $\Delta = \Delta_c$ with $|\beta_2| = |\beta_6|$ and $|\beta_5| = |\beta_1|$ satisfies $|\beta_i| \neq 1$, which is the reason of the conventional BBC breakdown since the topological phase transition always occurs at $|\beta_i| = 1$ in Hermitian systems. It can be seen in Fig.~\ref{fig: beta_Abs} that there are also crossings at $\Delta=-0.2$ and $\Delta=0$ with $|\beta|=1$, which are ensured by the TRS$^\dagger$ and are coincidentally the topological transition points under the PBC.

\section{Conclusions}
\label{sec: Conclusions}
In this paper, we systematically establish the bulk-boundary correspondence and give the topological invariants for various symmetry classes in non-Hermitian systems via the scattering theory. Compared to the prior GBZ-based approach, our method does not require computing the GBZ of the system which is usually cumbersome and time-consuming in practice. In addition, the numerical algorithm of our method only involving the matrix inversion rather than spectral decomposition, which causes amplified float-point errors due to the NHSE. Furthermore, in simple models, the topological invariants can be computed analytically by solving a generalized eigenproblem of $H(\beta)$, where $H(\beta)$ is the Bloch Hamiltonian $H(k)$ after substitution $e^{ik} \rightarrow \beta$.

When applying our method to calculate the topological invariants, there is a significant distinction between the non-Hermitian and the Hermitian cases. It is found that in some non-Hermitian systems, there is a discontinuity in the process of calculating the topological invariant, which results in a new type of topological phase transition and an unstable phase with topological boundary states, dubbed the critical topological phase transition and the critical topological phase. In contrast to ordinary topological phase transitions, the critical topological phase transition occurs without typical bulk energy gap closing. We find that in the prior GBZ viewpoint, it is due to a dramatic change of the GBZ. The change of topological invariant caused by the dramatic change of GBZ is usually difficult to characterize by directly calculating the GBZ, but our method provide an alternative way to detect it. It seems that the critical topological phase transition has the potential to design sensers with enhanced sensitivity against perturbations, which will be left for future study.

\begin{acknowledgements}
    The authors thank Yang Gao for discussions. This work was supported by the National Natural Science Foundation of China (12234017). .
\end{acknowledgements}

\appendix
\section{Mahaux-Weidenm{\"u}ller formula}
\label{sec: Mahaux}
\begin{figure}
    \includegraphics[width=0.9\columnwidth]{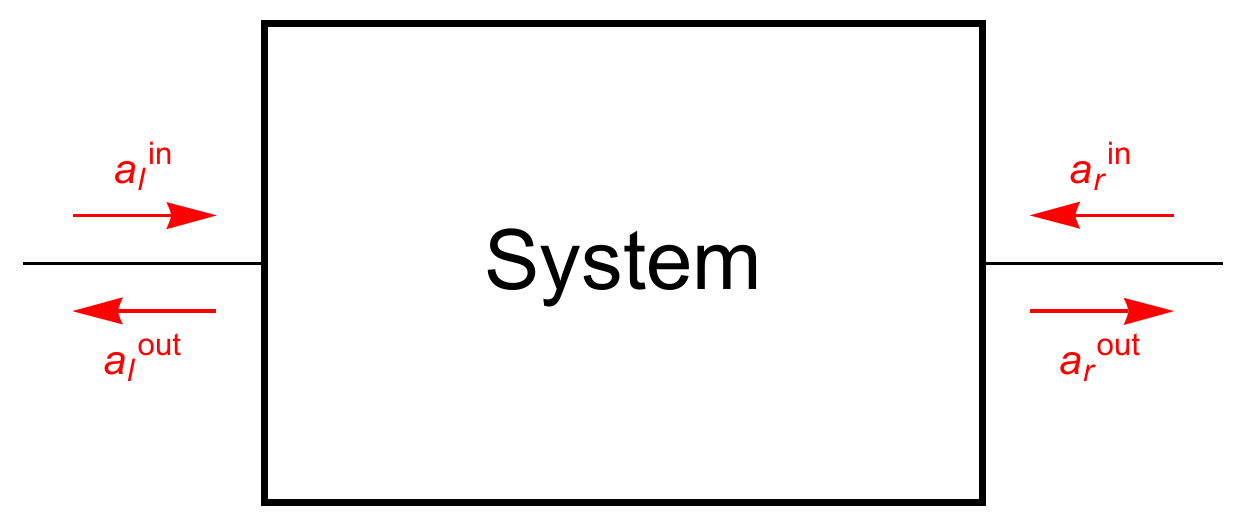}
    \caption{Schematic figure illustrating the input-output scattering formalism for the left and right channels case. The scattering medium has left (l) and right (r) optical channels and $\bm{a}^{\text{in(out)}}_{l,r}$ represents the the electromagnetic fields of the incoming (outgoing) modes at each optical channel.}
    \label{fig: scattering_LR}
\end{figure}

In this appendix, we derive the Mahaux-Weidenm{\"u}ller formula Eq.~(\ref{eq: Mahaux}). Assume the system is attached with left and right optical channels, or left and right leads as shown in Fig.~\ref{fig: scattering_LR}. The scattering matrix $S(\omega)$ gives the input-output relation for optical amplitudes at frequency $\omega$ \cite{doi:10.1080/00018732.2021.1876991}:
\begin{align}
    \bm{a}^{\text{out}} = S(\omega) \bm{a}^{\text{in}},
\end{align}
where $\bm{a}^{\text{in}}$ and $\bm{a}^{\text{out}}$ are the input and output optical amplitudes, respectively. In a system with left and right optical channels as shown in Fig.~\ref{fig: scattering_LR}, the amplitudes can be divided into two parts 
depending on different channels (denoted as the left channel and the right channel):
\begin{align}
    \bm{a}^{\text{in}} = \begin{pmatrix}
        \bm{a}^{\text{in}}_l \\
        \bm{a}^{\text{in}}_r
    \end{pmatrix}, \; \; \;
    \bm{a}^{\text{out}} = \begin{pmatrix}
        \bm{a}^{\text{out}}_l \\
        \bm{a}^{\text{out}}_r
    \end{pmatrix},
\end{align}
and the elements in $S(\omega)$ can be divided into four block
matrices as
\begin{align}
    S = \begin{pmatrix}
        r & t' \\
        t & r'
    \end{pmatrix}.
\end{align}

The Hamiltonian of the total system including left and right optical channels are \cite{Groth_2014}
\begin{align}
    H = \begin{pmatrix}
        \ddots & V^l_L & & & & & \\
        (V^l_L)^\dagger & H^l_L & V^l_L & & & & \\
         & (V^l_L)^\dagger & H^l_L & V^l_{\text{LS}} & & & \\
         &  & (V^l_{\text{LS}})^\dagger & H_S & V^r_{\text{LS}} & & \\
         & &  & (V^r_{\text{LS}})^\dagger & H^r_L & V^r_{L} & \\
         & & &  & (V^r_{L})^\dagger & H^r_L & V^r_{L} \\
         & & &  & & (V^r_{L})^\dagger & \ddots
    \end{pmatrix},
    \label{eq: totalH}
\end{align}
where $H_S$ is the Hamiltonian matrix of the scattering region $S$. $H^l_L$ ($H^r_L$) is the Hamiltonian of one unit cell of the left (right) optical channel, while the block submatrix $V^l_L$ ($V^r_L$) is the Hamiltonian connecting one unit cell of the left (right) optical channel to the next. Finally, $V^l_{\text{LS}}$ ($V^r_{\text{LS}}$) is the hopping from the system to the left (right) optical channel. By translational invariance of the optical channel, The eigenstates of the translation operator in the optical channel take the form
\begin{align}
    \phi_n^{l(r)}(j) = (\lambda_n^{l(r)})^j u_n^{l(r)},
\end{align}
such that they obey the Schr{\"o}dinger equation
\begin{align}
    \left[H_L^{l(r)}+V_L^{l(r)}+(V^{l(r)})^\dagger \right] u_n^{l(r)} = \omega u_n^{l(r)},
    \label{eq: lead_eq}
\end{align}
where $u_n^{l(r)}$ is the n-th eigenvector in the left (right) optical channel, and $\lambda_n^{l(r)}$ the n-th eigenvalue in the left (right) optical channel.

By solving the eigenproblem of the total Hamiltonian (\ref{eq: totalH}), we obtain the following equation
\begin{align}
    & \begin{pmatrix}
        -V^l_L U^l_{\text{out}} & V^l_{\text{LS}} & 0 \\
        (V^l_{\text{LS}})^\dagger U^l_{\text{out}} (\Lambda^l_{\text{out}})^{-1} & H_S-\omega & V^r_{\text{LS}} U^r_{\text{out}} \Lambda^r_{\text{out}} \\
        0 & (V^r_{\text{LS}})^\dagger & -(V^r_L)^\dagger U^r_{\text{out}}
    \end{pmatrix}
    \begin{pmatrix}
        r \\ \psi_S \\ t
    \end{pmatrix} \notag \\
    = & \begin{pmatrix}
        V^l_L U^l_{\text{in}} \\
        -(V^l_{\text{LS}})^\dagger U^l_{\text{in}} (\Lambda^l_{\text{in}})^{-1} \\
        0
    \end{pmatrix},
    \label{eq: S_eq}
\end{align}
with $U^{l(r)}_{\text{in}}$ and $U^{l(r)}_{\text{out}}$ wave functions of incoming and outgoing modes at the left (right) optical channel, and $\Lambda=\text{diag}(\lambda_i)$. $r$ is the reflection matrix, $t$ is the transmission matrix, and $\psi_S$ is a matrix whose column vectors are wave functions in the scattering region. We set $V^{l(r)}_L=w \mathbb{I}$ and $H^{l(r)}_L=0$. It can be solved from Eq.~(\ref{eq: lead_eq}) that $U^{l(r)}_{\text{in,out}} \propto \mathbb{I}$, and we adopt the sign convention that $U^{l}_{\text{in,out}}=\mathbb{I}$ and $U^{r}_{\text{in,out}}=-\mathbb{I}$. If the frequency $\omega$ of the signal is in the energy gap of the system and the length of the scattering region is sufficiently large, the transmission process is negligible, and the reflection matrix can be solved from Eq.~(\ref{eq: S_eq}) as 
\begin{align}
    r(\omega) & = \left[\mathbb{I} - i V_{\text{LS}} G(\omega) V^{\dagger}_{\text{LS}} \right]^{-1} \left[ \mathbb{I} + i V_{\text{LS}} G(\omega) V^{\dagger}_{\text{LS}} \right] \notag \\ 
    & \equiv \frac{\mathbb{I} + i V_{\text{LS}} G(\omega) V^{\dagger}_{\text{LS}}}{\mathbb{I} - i V_{\text{LS}} G(\omega) V^{\dagger}_{\text{LS}}},
\end{align}
where $G(\omega) = (\omega-H_S)^{-1}$ is the Green's function of the system after dropping optical channels.

\section{Topological invariant in terms of the Green's function}
\label{sec: Topo_inv_G}
In this Appendix, we express the topological invariants protected by different symmetry in terms of the Green's function.
By the spectral decomposition of the Green's function, after taking the thermodynamics limit, the diagonal block of the real space Green's function at the first unit cell can be written as 
\begin{align}
    G_{11}(\omega) = \frac{A}{\omega} + (\text{regular terms at }\omega=0),
    \label{eq: G11}
\end{align}
where $G_{11}(\omega)$ and $A$ are $M \times M$ matrices when the number of the internal degree of freedom is $M$ at each unit cell of the system.

Consider the system with the sublattice symmetry.
Due to $\Gamma \, G(\omega) \, \Gamma^{-1} = -G(-\omega)$, the commutator $\left[ A , \Gamma \right] =0$, which implies that $A$ and $\Gamma$ can be diagonalized simultaneously. Assume that in the diagonalized basis,
\begin{align}
    \Gamma & = \text{diag}(\mathbb{I}_{M/2 \times M/2}, -\mathbb{I}_{M/2 \times M/2}), \notag \\
    A & = \text{diag}(A_{1,+},\ldots,A_{p,+},0, \notag \\
        & \ldots,A_{1,-},\ldots,A_{q,-},0,\ldots). 
    \label{eq: A}
\end{align}
For simplicity, we let the coupling between the system and the fictitious left optical channel be $V_{\text{LS}}=\mathbb{I}_{M\times M}$. By Eq.~(\ref{eq: r_G}) and Eq.~(\ref{eq: A}), 
    \begin{align}
        & \lim_{\omega \rightarrow 0} \Gamma r(\omega) \notag \\
        = & \text{diag}\left(-\mathbb{I}_{p \times p},\mathbb{I}_{(M/2-p) \times (M/2-p)},\mathbb{I}_{q \times q},-\mathbb{I}_{(M/2-q)\times (M/2-q)}\right),
    \end{align}
which implies that 
\begin{align}
    \lim_{\omega \rightarrow 0} -\frac{\Tr (\Gamma r(\omega))}{2}  & = p-q \notag \\
    & = \text{rank}(A \, \Pi_+) - \text{rank}(A \, \Pi_-),
    \label{eq: Topo_A}
\end{align}
where $\text{rank}(\cdot)$ takes the matrix rank and $\Pi_\pm$ is the projection operator to the positive (negative) chirality subspace.
Therefore, the topological invariant protected by the sublattice symmetry is encoded in the coefficient matrix of the $\frac{1}{\omega}$ term of $G_{11}(\omega)$. In Sec.~\ref{sec: BBC_SLS}, Eq.~(\ref{eq: Topo_A}) will be used to build the BBC in Eq.~(\ref{eq: SLS_BBC}).

Consider the system with an additional TRS$^\dagger$ symmetry. Assume that in the diagonalized basis 
\begin{align}
    \Gamma & = \text{diag}(\mathbb{I}_{M/2 \times M/2}, -\mathbb{I}_{M/2 \times M/2}), \notag \\
    \mathcal{U}_C & = \begin{pmatrix}
        0 & \mathbb{I}_{M/2 \times M/2} \\
        \mathbb{I}_{M/2 \times M/2} & 0
    \end{pmatrix}, \notag \\
    A & = \text{diag}(A_{1,+},\ldots,A_{p,+},0, \notag \\
        & \ldots,A_{1,-},\ldots,A_{q,-},0,\ldots).
    \label{eq: TC_diag}
\end{align}
By Eq.~(\ref{eq: TC_G}), $\mathcal{U}_C A \mathcal{U}_C^{-1} = A^T$, which requires $p=q$ and $A_{+,i}=A_{-,i}, \; i=1,\ldots,p$ in Eq.~(\ref{eq: TC_diag}). It means that non-zero eigenvalues of $A$ form Kramer pairs. By Eq.~(\ref{eq: TC_diag}), 
\begin{widetext}
\begin{align}
    & \lim_{\omega \rightarrow 0} i \mathcal{V}_C^{\dagger} \Gamma r(\omega) \mathcal{V}_C \notag \\
    = & \begin{pmatrix}
        0 & \text{diag}(-\mathbb{I}_{p \times p}, \mathbb{I}_{(M/2-p) \times (M/2-p)}) \\
        \text{diag}(\mathbb{I}_{p \times p}, -\mathbb{I}_{(M/2-p) \times (M/2-p)}) & 0
    \end{pmatrix}.
\end{align}
\end{widetext}
It follows that 
\begin{align}
    (-1)^p = (-1)^{\frac{M}{2}\left( \frac{M}{2} -1 \right)/2} \text{Pf}(i \mathcal{V}_C^{\dagger} \Gamma r(0) \mathcal{V}_C),
    \label{eq: Topo_A_TRS}
\end{align}
where $p$ is the number of Kramer pairs of non-zero eigenvalues of $A$. In Sec.~\ref{sec: BBC_TRS}, Eq.~(\ref{eq: Topo_A_TRS}) will be used to build the BBC in Eq.~(\ref{eq: TRS_BBC}).

\section{Green's function in terms of generalized eigenvalues and eigenvectors}
\label{sec: Green_app}
In this Appendix, we give the expression of $G_{11}(\omega)$ by solving a generalized eigenproblem. This method is originally given in Ref.~\cite{PhysRevB.95.235143}, and some changes are made since now the non-Hermitian cases are considered. Our result is an exact formula for the particular matrix block $G_{11}(\omega)$ of the Green's function in the non-Hermitian mutiband system, which is a supplementary result of previous results of Green's function in the non-Hermitian single-band system \cite{PhysRevB.103.L241408, PhysRevB.105.045122}.

As in Sec.~\ref{sec: analy} of the main text, we consider the following tight-binding model
\begin{align}
    H = \sum_{n,\mu,\nu} c^{\dagger}_{n,\mu} h_{\mu,\nu} c_{n,\nu} + c^{\dagger}_{n+1,\mu} V_{\mu,\nu} c_{n,\nu} + c^{\dagger}_{n,\mu} W_{\mu,\nu} c_{n+1,\nu},
    \label{eq: Ham}
\end{align}
where $c^{\dagger}_{n,\mu}$ ($c_{n,\mu}$) is a creation (an annihilation) operator of a particle with index $\mu$ (we assume a unit cell is composed of $q$ degrees of freedom, hence, $\mu=1,\ldots,M$) in the $n$-th unit cell.  

Consider the following eigenequation in real space
\begin{align}
    (\omega - H) \psi = 0
\end{align}
which can be written in components as 
\begin{align}
    g^{-1}(\omega) \psi(n) - V \psi(n-1) - W \psi(n+1) = \mathbf{0},
\end{align}
where $g^{-1}(\omega) = \omega - h$. Furthermore, we impose the open boundary conditions $\psi(0) = \mathbf{0} = \psi(N+1)$. Assume that $V$ is not singular, the two-components quantities $\Psi(n) = \left[ \psi(n-1)^T, \psi(n)^T \right]^T$ satisfies 
$\Psi(n-1) = T(\omega) \Psi(n)$, where the matrix
\begin{align}
    T(\omega) = \begin{pmatrix}
        V^{-1} g^{-1}(\omega) & -V^{-1} W \\
        \mathbb{I} & 0
    \end{pmatrix}
    \label{eq: left_transfer}
\end{align}
is the transfer matrix.

Denote $G_{11,N}(\omega)$ be the diagonal element of the Green's function at the first unit cell in a system with $N$ unit cells. By the Dyson equation, 
\begin{align}
    (g^{-1}(\omega) - W G_{11,N-1}(\omega) V) G_{11,N}(\omega) = \mathbb{I},
    \label{eq: Dyson}
\end{align}
where $g(\omega)=\frac{1}{\omega-h}$ is the Green's function of the unit cell after dropping all inter-cell couplings. Let $X_n(\omega) = G_{11,n}(\omega) V$, Eq.~(\ref{eq: Dyson}) is equivalent to 
\begin{align}
    V^{-1} g^{-1}(\omega) X_N(\omega) - V^{-1} W X_{N-1}(\omega) X_{N}(\omega) = \mathbb{I}
\end{align}
or 
\begin{align}
    \begin{pmatrix}
        \mathbb{I} \\
        X_N(\omega)
    \end{pmatrix} = T(\omega) 
    \begin{pmatrix}
        \mathbb{I} \\
        X_{N-1}(\omega)
    \end{pmatrix} X_N(\omega),
    \label{eq: Dyson_M}
\end{align}
where $T(\omega)$ is the transfer matrix of $\omega-H$ as in Eq.~(\ref{eq: left_transfer}).
Eq.~(\ref{eq: Dyson_M}) can be applied iteratively to obtain 
\begin{align}
    \begin{pmatrix}
        \mathbb{I} \\
        X_N(\omega)
    \end{pmatrix} = T(\omega)^N 
    \begin{pmatrix}
        \mathbb{I} \\
        0
    \end{pmatrix} X_1(\omega) \cdots X_N(\omega),
\end{align}
which implies the final result 
\begin{align}
    G_{11,N}(\omega) = [T(\omega)^N]_{21} [T(\omega)^N]_{11}^{-1} V^{-1},
\end{align}
where $T(\omega)^N$ is divided into four square matrix blocks with the same size and $[T(\omega)^N]_{\mu \nu}$ denote the block of $T(\omega)^N$ at the $\mu$-th row and the $\nu$-th column.
Therefore, in a one-dimensional system with the above Hamiltonian and $N$ unit cells, 
\begin{align}
    G_{11}(\omega) = [T(\omega)^N]_{21} [T(\omega)^N]_{11}^{-1} V^{-1}.
    \label{eq: G_11}
\end{align}

Assume that $T(\omega)$ can be diagonalized as 
\begin{align}
    \begin{pmatrix}
        U_{11} & U_{12} \\
        U_{21} & U_{22}
    \end{pmatrix}^{-1} T(\omega) 
    \begin{pmatrix}
        U_{11} & U_{12} \\
        U_{21} & U_{22}
    \end{pmatrix} = 
    \begin{pmatrix}
        B_1 & 0 \\
        0 & B_2
    \end{pmatrix},
\end{align}
where $B_1 = \text{diag}(\beta_1, \ldots, \beta_M)$, $B_2 =  \text{diag}(\beta_{M+1}, \ldots, \beta_{2M})$. By Eq.~(\ref{eq: G_11}),
\begin{align}
    G_{11}(\omega) = & \left[ U_{21} B_1^N (U^{-1})_{11} + U_{22} B_2^N (U^{-1})_{21} \right] \notag \\
    & \left[ U_{11} B_1^N (U^{-1})_{11} + U_{12} B_2^N (U^{-1})_{21} \right]^{-1} V^{-1}.
    \label{eq: G11_by_parts}
\end{align}
If $U_{11}$ and $(U^{-1})_{11}$ are non-singular, 
\begin{align}
    G_{11}(\omega) = & \left[ U_{21} + U_{22} B_2^N (U^{-1})_{21} (U^{-1})_{11}^{-1} B_1^{-N} \right] \notag \\
    & \left[ U_{11} + U_{12} B_2^{N} (U^{-1})_{21} (U^{-1})_{11}^{-1} B_1^{-N} \right]^{-1} V^{-1}.
    \label{eq: G11_UB}
\end{align}
If $\beta_1,\ldots,\beta_{2M}$ are ordered to satisfy $|\beta_1| \geqslant |\beta_2| \geqslant \ldots \geqslant |\beta_{M}| > |\beta_{M+1}| \geqslant \ldots \geqslant |\beta_{2M}|$, in the thermodynamics limit,
\begin{align}
    \lim_{N \rightarrow \infty} G_{11}(\omega) = U_{21} U_{11}^{-1} V^{-1},
    \label{eq: G11_U}
\end{align}
since terms in Eq.~(\ref{eq: G11_by_parts}) involving $U_{21}$ and $U_{11}$ are dominant comparing to the other terms.

In contrast to Hermitian cases, conditions that $U_{11}$ and $(U^{-1})_{11}$ are non-singular and $|\beta_1| \geqslant |\beta_2| \geqslant \ldots \geqslant |\beta_{M}| > |\beta_{M+1}| \geqslant \ldots \geqslant |\beta_{2M}|$ may not be achievable simultaneously in some non-Hermitian models with some discrete symmetries. However, by a suitable way of dividing eigenvectors of $T(\omega)$ into two groups, conditions that $U_{11}$ and $(U^{-1})_{11}$ are non-singular and terms in Eq.~(\ref{eq: G11_by_parts}) involving $U_{21}$ and $U_{11}$ are dominant comparing to the other terms can be achieved simultaneously without requiring $|\beta_1| \geqslant |\beta_2| \geqslant \ldots \geqslant |\beta_{M}| > |\beta_{M+1}| \geqslant \ldots \geqslant |\beta_{2M}|$, and Eq.~(\ref{eq: G11_U}) is still valid as a consequence. 
We discuss a four-band system in Sec.~\ref{sec: critical_topo} and a TRS$^{\dagger}$ protected system in Sec.~\ref{sec: TRS_analy} as two examples.

We denote $H(\beta) = h+V\beta+W \beta^{-1}$. The eigenvector of $T(\omega)$ with eigenvalue $\beta$ has the form $(\beta x^T , x^T)^T$ with the column vector $x$ being the nullvector of $\omega - H(\beta)$, i.e., $[\omega - H(\beta)] x = 0$. By using this fact,
\begin{align}
    \begin{pmatrix}
        U_{11} \\
        U_{21} 
    \end{pmatrix} = 
    \begin{pmatrix}
        \beta_1 x_1 , \ldots, \beta_M x_M \\
        x_1, \ldots, x_M
    \end{pmatrix}.
\end{align}
Eq.~(\ref{eq: G11_U}) becomes
\begin{align}
    & \lim_{N \rightarrow \infty} G_{11}(\omega) \notag \\
    = & (x_1,\ldots,x_M)(\omega) \text{diag}(\beta_1^{-1},\ldots,\beta_M^{-1}) (x_1,\ldots,x_M)^{-1}(\omega) ,
    \label{eq: G11_therm}
\end{align}
where $\beta_1(\omega),\ldots,\beta_M(\omega)$ are $M$ selected $\beta$ solutions of $\det[\omega - H(\beta)]= 0$ and $x_i(\omega),i=1,\ldots,M$ are corresponding nullvectors of $[\omega - H(\beta_i)],i=1,\ldots,M$. For general systems without any discrete symmetry, $\beta_1(\omega),\ldots,\beta_M(\omega)$ are $M$ $\beta$ solutions of $\det[\omega - H(\beta)]= 0$ with largest magnitudes. 

Before the end of this Appendix, we want to further prove that $x_i(\omega)$ depends linearly on $\omega$ as its first order perturbation. Since $T(\omega)=T(0)+\omega \delta T$, treat $\omega \delta T$ as perturbation, the first order perturbation of the eigenvector $v_i(\omega)=v_i(0)+\omega \delta v_i$ of $T(\omega)$ is 
\begin{align}
    \delta v_i = \sum_{j \neq i} \frac{w_j^T(0) (\omega \delta T) v_i(0)}{w_j^T(0) v_j(0) \left[\beta_i(0) - \beta_j(0)\right]} v_j(0),
\end{align}
where $v_j(0)$ ($w_j^T(0)$) is the right (left) eigenvector of $T(0)$ with eigenvalue $\beta_j(0)$. It follows that the first order correction of $(x_1,\ldots,x_M)(\omega) = (x_1(0)+\omega \delta x_1,\ldots,x_M(0)+\omega \delta x_M)$ depends linearly on $\omega$.

\section{Nullvector selection in a four-band model}
In this Appendix, we explain the nullvector selection of $x_1(\omega),\ldots,x_M(\omega)$ in Sec.~\ref{sec: critical_topo} of the main text, which illustrates the case that conditions that $U_{11}$ and $(U^{-1})_{11}$ are non-singular and $|\beta_1| \geqslant |\beta_2| \geqslant \ldots \geqslant |\beta_{M}| > |\beta_{M+1}| \geqslant \ldots \geqslant |\beta_{2M}|$ are not achievable simultaneously in Eq.~(\ref{eq: G11_UB}). In this example, the Hamiltonian $H(\beta) = \begin{pmatrix}
    0 & H_-(\beta) \\
    H_+(\beta) & 0
\end{pmatrix}$ where 
\begin{align}
    & H_{\pm}(\beta) \notag \\
    = & \begin{pmatrix}
        a_{\pm} \frac{(\beta-\beta_{\pm,a,1})(\beta-\beta_{\pm,a,2})}{\beta} & c \\
        c & b_{\pm} \frac{(\beta-\beta_{\pm,b,1})(\beta-\beta_{\pm,b,2})}{\beta}
    \end{pmatrix}.
\end{align}

When the off-diagonal coupling $c$ of $H_{\pm}(\beta)$ is 0, the Hamiltonian is the direct sum of two independent two-band Hamiltonians with the sublattice symmetry $H=H_a \oplus H_b$, where the index $a$ represents the first diagonal element of $H_{\pm}$ and the index $b$ represents the second diagonal element of $H_{\pm}$. The Green's function can be decomposited similarly as 
\begin{align}
    G_{11}(\omega) = G_{11,a}(\omega) \oplus G_{11,b}(\omega).
    \label{eq: G11_dec}
\end{align}
Since $|\beta_{+,a,1}|,|\beta_{+,a,2}|>|\beta_{-,a,1}|,|\beta_{-,a,2}|>|\beta_{+,b,1}|,|\beta_{+,b,2}|>|\beta_{-,b,1}|,|\beta_{-,b,2}|$. For a small frequency $\omega \neq 0$, similar ordering $|\beta_{+,a,1}(\omega)|,|\beta_{+,a,2}(\omega)|>|\beta_{-,a,1}(\omega)|,|\beta_{-,a,2}(\omega)|>|\beta_{+,b,1}(\omega)|,|\beta_{+,b,2}(\omega)|>|\beta_{-,b,1}(\omega)|,|\beta_{-,b,2}(\omega)|$ is used to label the $\beta$ solutions of the equation $\det \left[\omega-H(\beta)\right]=0$. Notice that $\beta_{+,a,1}(\omega),\beta_{+,a,2}(\omega),\beta_{-,a,1}(\omega),\beta_{-,a,2}(\omega)$ are four $\beta$ solutions with the largest magnitude, and the corresponding nullvectors of them are linear dependent. Therefore, conditions that $U_{11}$ and $(U^{-1})_{11}$ are non-singular and $|\beta_1| \geqslant |\beta_2| \geqslant \ldots \geqslant |\beta_{M}| > |\beta_{M+1}| \geqslant \ldots \geqslant |\beta_{2M}|$ are not achievable simultaneously in this model, and Eq.~(\ref{eq: G11_U}) is invalid when $U_{21}$ is taken to be the matrix composed by the corresponding nullvectors of $\beta_{+,a,1}(\omega),\beta_{+,a,2}(\omega),\beta_{-,a,1}(\omega),\beta_{-,a,2}(\omega)$. However, Eq.~(\ref{eq: G11_dec}) suggests that $\beta_{\pm,a,1}(\omega),\beta_{\pm,a,2}(\omega)$ and $\beta_{\pm,b,1}(\omega),\beta_{\pm,b,2}(\omega)$ should be treated independently. In Eq.~(\ref{eq: G11_by_parts}), terms involving them are independent. Since $\beta_{+,a,1}(\omega),\beta_{+,a,2}(\omega)$ are two $\beta$ solutions with the largest magnitude from $\beta_{\pm,a,1}(\omega),\beta_{\pm,a,2}(\omega)$ and $\beta_{+,b,1}(\omega),\beta_{+,b,2}(\omega)$ are two $\beta$ solutions with the largest magnitude from $\beta_{\pm,b,1}(\omega),\beta_{\pm,b,2}(\omega)$, if $U_{21}$ is taken to be the matrix composed by the corresponding nullvectors of $\beta_{+,a,1}(\omega),\beta_{+,a,2}(\omega),\beta_{+,b,1}(\omega),\beta_{+,b,2}(\omega)$, Eq.~(\ref{eq: G11_U}) is still valid. As a consequence, in Eq.~(\ref{eq: G11_therm}), $\beta_1(\omega),\ldots,\beta_M(\omega)$ should be $\beta_{+,a,1}(\omega),\beta_{+,a,2}(\omega),\beta_{+,b,1}(\omega),\beta_{+,b,2}(\omega)$, and $x_1(\omega),\ldots,x_M(\omega)$ should be nullvectors corresponding to $\beta_{+,a,1}(\omega),\beta_{+,a,2}(\omega),\beta_{+,b,1}(\omega),\beta_{+,b,2}(\omega)$.

When $c \neq 0$, the $\beta$ solutions of $\det[\omega-H(\beta)]=0$ deviate slightly from $\beta_{\pm,a,i}(\omega)$ and $\beta_{\pm,b,i}(\omega)$ while the order of their magnitudes keeps invariant, therefore, we denote them by $\beta'_{\pm,a,i}(\omega)$ and $\beta'_{\pm,b,i}(\omega)$ with the same subscript indices. In contrast to the above $c=0$ case, in Eq.~(\ref{eq: G11_by_parts}), terms involving $\beta'_{\pm,a,1}(\omega),\beta'_{\pm,a,2}(\omega)$ and $\beta'_{\pm,b,1}(\omega),\beta'_{\pm,b,2}(\omega)$ are not independent. Therefore, in Eq.~(\ref{eq: G11_therm}), $\beta_1(\omega),\ldots,\beta_M(\omega)$ should be $\beta'_{+,a,1}(\omega),\beta'_{+,a,2}(\omega),\beta'_{-,a,1}(\omega),\beta'_{-,a,2}(\omega)$, and $x_1(\omega),\ldots,x_M(\omega)$ should be nullvectors corresponding to $\beta'_{+,a,1}(\omega),\beta'_{+,a,2}(\omega),\beta'_{-,a,1}(\omega),\beta'_{-,a,2}(\omega)$.

\section{Hermitian topological phase near $c=0$}
\label{sec: Herm_c}
In this appendix, we prove that the new type of topological phase transition discussed in Sec.~\ref{sec: critical_topo} does not exist in Hermitian systems, in other words, we prove that the discontinuity of $(x_1,\ldots,x_M)$ and $(\beta_1,\ldots,\beta_M)$ selection at $c=0$ does not exist in Hermitian cases.

Denote 
\begin{align}
    H(\beta) = \begin{pmatrix}
    0 & h_- + V_- \beta + W_- \beta^{-1} \\
    h_+ + V_+ \beta + W_+ \beta^{-1} & 0
\end{pmatrix}.
\end{align}
Since the Hamiltonian is Hermitian, 
\begin{align}
    h_+=h^\dagger_- \; , \; V_+ = W_-^\dagger \; , \; W_+=V_-^\dagger,
\end{align}
and it follows that $H_+(\beta)= h_+ + V_+ \beta + W_+ \beta^{-1}$ and $H_-(\beta)=h_- + V_- \beta + W_- \beta^{-1}$ satisfy $H_+(\beta) = H_-(\frac{1}{\beta^*})^\dagger$. Hence, $\beta$ solutions of $0=\det [H(\beta)] = (-1)^{\frac{M}{2}(\frac{M}{2}+1)} \det[H_+(\beta)] \det[H_-(\beta)]$ come into $\beta$ and $\frac{1}{\beta^*}$ pairs. Hence, $\beta$ solutions of $0=\det [H(\beta)]$ ordered by their magnitudes must satisfy $|\beta_1| \geqslant |\beta_2| \geqslant \ldots \geqslant |\beta_{M}| > 1 > |\beta_{M+1}| \geqslant \ldots \geqslant |\beta_{2M}|$, i.e., magnitudes of $\beta_1,\ldots,\beta_M$ and $\beta_{M+1},\ldots,\beta_{2M}$ are separated by 1. Therefore, in Hermitian systems, $\beta_1,\ldots,\beta_M$ appearing in Eq.~(\ref{eq: G11_therm}) are $\beta$ solutions of $0=\det [H(\beta)]$ whose magnitudes are larger than 1. By our discussion in the main text, at $c=0$, $\beta_1,\ldots,\beta_M$ are determined by two independent sectors,
\begin{align}
    \{ \beta_1, \ldots, \beta_M \} = \{ \beta_{i,a} | \; |\beta_{i,a}|>1 \} \cup \{ \beta_{i,b} | \; |\beta_{i,b}|>1 \}.
\end{align}
Due to 
\begin{align}
    \{ \beta_i | \; |\beta_i|>1 \} = \{ \beta_{i,a} | \; |\beta_{i,a}|>1 \} \cup \{ \beta_{i,b} | \; |\beta_{i,b}|>1 \},
\end{align}
where $\{ \beta_i | \; |\beta_i|>1 \}$ represents $\beta$ solutions without $a,b$ section labels,
there is no discontinuity of $(x_1,\ldots,x_M)$ and $(\beta_1,\ldots,\beta_M)$ selection in Hermitian systems. Therefore, in Hermitian systems, the topological invariant and the number of localized zero energy modes are always the same for $c=0$ and $c \neq 0$ ($c$ is small) phases.

\bibliography{Ref}

\begin{thebibliography}{45}%
\makeatletter
\providecommand \@ifxundefined [1]{%
 \@ifx{#1\undefined}
}%
\providecommand \@ifnum [1]{%
 \ifnum #1\expandafter \@firstoftwo
 \else \expandafter \@secondoftwo
 \fi
}%
\providecommand \@ifx [1]{%
 \ifx #1\expandafter \@firstoftwo
 \else \expandafter \@secondoftwo
 \fi
}%
\providecommand \natexlab [1]{#1}%
\providecommand \enquote  [1]{``#1''}%
\providecommand \bibnamefont  [1]{#1}%
\providecommand \bibfnamefont [1]{#1}%
\providecommand \citenamefont [1]{#1}%
\providecommand \href@noop [0]{\@secondoftwo}%
\providecommand \href [0]{\begingroup \@sanitize@url \@href}%
\providecommand \@href[1]{\@@startlink{#1}\@@href}%
\providecommand \@@href[1]{\endgroup#1\@@endlink}%
\providecommand \@sanitize@url [0]{\catcode `\\12\catcode `\$12\catcode
  `\&12\catcode `\#12\catcode `\^12\catcode `\_12\catcode `\%12\relax}%
\providecommand \@@startlink[1]{}%
\providecommand \@@endlink[0]{}%
\providecommand \url  [0]{\begingroup\@sanitize@url \@url }%
\providecommand \@url [1]{\endgroup\@href {#1}{\urlprefix }}%
\providecommand \urlprefix  [0]{URL }%
\providecommand \Eprint [0]{\href }%
\providecommand \doibase [0]{http://dx.doi.org/}%
\providecommand \selectlanguage [0]{\@gobble}%
\providecommand \bibinfo  [0]{\@secondoftwo}%
\providecommand \bibfield  [0]{\@secondoftwo}%
\providecommand \translation [1]{[#1]}%
\providecommand \BibitemOpen [0]{}%
\providecommand \bibitemStop [0]{}%
\providecommand \bibitemNoStop [0]{.\EOS\space}%
\providecommand \EOS [0]{\spacefactor3000\relax}%
\providecommand \BibitemShut  [1]{\csname bibitem#1\endcsname}%
\let\auto@bib@innerbib\@empty
\bibitem [{\citenamefont {Thouless}\ \emph {et~al.}(1982)\citenamefont
  {Thouless}, \citenamefont {Kohmoto}, \citenamefont {Nightingale},\ and\
  \citenamefont {den Nijs}}]{PhysRevLett.49.405}%
  \BibitemOpen
  \bibfield  {author} {\bibinfo {author} {\bibfnamefont {D.~J.}\ \bibnamefont
  {Thouless}}, \bibinfo {author} {\bibfnamefont {M.}~\bibnamefont {Kohmoto}},
  \bibinfo {author} {\bibfnamefont {M.~P.}\ \bibnamefont {Nightingale}}, \ and\
  \bibinfo {author} {\bibfnamefont {M.}~\bibnamefont {den Nijs}},\ }\bibfield
  {title} {\enquote {\bibinfo {title} {Quantized hall conductance in a
  two-dimensional periodic potential},}\ }\href {\doibase
  10.1103/PhysRevLett.49.405} {\bibfield  {journal} {\bibinfo  {journal} {Phys.
  Rev. Lett.}\ }\textbf {\bibinfo {volume} {49}},\ \bibinfo {pages} {405--408}
  (\bibinfo {year} {1982})}\BibitemShut {NoStop}%
\bibitem [{\citenamefont {Wen}\ and\ \citenamefont
  {Niu}(1990)}]{PhysRevB.41.9377}%
  \BibitemOpen
  \bibfield  {author} {\bibinfo {author} {\bibfnamefont {X.~G.}\ \bibnamefont
  {Wen}}\ and\ \bibinfo {author} {\bibfnamefont {Q.}~\bibnamefont {Niu}},\
  }\bibfield  {title} {\enquote {\bibinfo {title} {Ground-state degeneracy of
  the fractional quantum hall states in the presence of a random potential and
  on high-genus riemann surfaces},}\ }\href {\doibase 10.1103/PhysRevB.41.9377}
  {\bibfield  {journal} {\bibinfo  {journal} {Phys. Rev. B}\ }\textbf {\bibinfo
  {volume} {41}},\ \bibinfo {pages} {9377--9396} (\bibinfo {year}
  {1990})}\BibitemShut {NoStop}%
\bibitem [{\citenamefont {Hasan}\ and\ \citenamefont
  {Kane}(2010)}]{RevModPhys.82.3045}%
  \BibitemOpen
  \bibfield  {author} {\bibinfo {author} {\bibfnamefont {M.~Z.}\ \bibnamefont
  {Hasan}}\ and\ \bibinfo {author} {\bibfnamefont {C.~L.}\ \bibnamefont
  {Kane}},\ }\bibfield  {title} {\enquote {\bibinfo {title} {Colloquium:
  Topological insulators},}\ }\href {\doibase 10.1103/RevModPhys.82.3045}
  {\bibfield  {journal} {\bibinfo  {journal} {Rev. Mod. Phys.}\ }\textbf
  {\bibinfo {volume} {82}},\ \bibinfo {pages} {3045--3067} (\bibinfo {year}
  {2010})}\BibitemShut {NoStop}%
\bibitem [{\citenamefont {Bernevig}\ \emph {et~al.}(2006)\citenamefont
  {Bernevig}, \citenamefont {Hughes},\ and\ \citenamefont
  {Zhang}}]{science.1133734}%
  \BibitemOpen
  \bibfield  {author} {\bibinfo {author} {\bibfnamefont {B.~Andrei}\
  \bibnamefont {Bernevig}}, \bibinfo {author} {\bibfnamefont {Taylor~L.}\
  \bibnamefont {Hughes}}, \ and\ \bibinfo {author} {\bibfnamefont {Shou-Cheng}\
  \bibnamefont {Zhang}},\ }\bibfield  {title} {\enquote {\bibinfo {title}
  {Quantum spin hall effect and topological phase transition in hgte quantum
  wells},}\ }\href {\doibase 10.1126/science.1133734} {\bibfield  {journal}
  {\bibinfo  {journal} {Science}\ }\textbf {\bibinfo {volume} {314}},\ \bibinfo
  {pages} {1757--1761} (\bibinfo {year} {2006})}\BibitemShut {NoStop}%
\bibitem [{\citenamefont {Kane}\ and\ \citenamefont
  {Mele}(2005{\natexlab{a}})}]{PhysRevLett.95.226801}%
  \BibitemOpen
  \bibfield  {author} {\bibinfo {author} {\bibfnamefont {C.~L.}\ \bibnamefont
  {Kane}}\ and\ \bibinfo {author} {\bibfnamefont {E.~J.}\ \bibnamefont
  {Mele}},\ }\bibfield  {title} {\enquote {\bibinfo {title} {Quantum spin hall
  effect in graphene},}\ }\href {\doibase 10.1103/PhysRevLett.95.226801}
  {\bibfield  {journal} {\bibinfo  {journal} {Phys. Rev. Lett.}\ }\textbf
  {\bibinfo {volume} {95}},\ \bibinfo {pages} {226801} (\bibinfo {year}
  {2005}{\natexlab{a}})}\BibitemShut {NoStop}%
\bibitem [{\citenamefont {Kane}\ and\ \citenamefont
  {Mele}(2005{\natexlab{b}})}]{PhysRevLett.95.146802}%
  \BibitemOpen
  \bibfield  {author} {\bibinfo {author} {\bibfnamefont {C.~L.}\ \bibnamefont
  {Kane}}\ and\ \bibinfo {author} {\bibfnamefont {E.~J.}\ \bibnamefont
  {Mele}},\ }\bibfield  {title} {\enquote {\bibinfo {title} {${Z}_{2}$
  topological order and the quantum spin hall effect},}\ }\href {\doibase
  10.1103/PhysRevLett.95.146802} {\bibfield  {journal} {\bibinfo  {journal}
  {Phys. Rev. Lett.}\ }\textbf {\bibinfo {volume} {95}},\ \bibinfo {pages}
  {146802} (\bibinfo {year} {2005}{\natexlab{b}})}\BibitemShut {NoStop}%
\bibitem [{\citenamefont {Fu}\ and\ \citenamefont
  {Kane}(2006)}]{PhysRevB.74.195312}%
  \BibitemOpen
  \bibfield  {author} {\bibinfo {author} {\bibfnamefont {Liang}\ \bibnamefont
  {Fu}}\ and\ \bibinfo {author} {\bibfnamefont {C.~L.}\ \bibnamefont {Kane}},\
  }\bibfield  {title} {\enquote {\bibinfo {title} {Time reversal polarization
  and a ${Z}_{2}$ adiabatic spin pump},}\ }\href {\doibase
  10.1103/PhysRevB.74.195312} {\bibfield  {journal} {\bibinfo  {journal} {Phys.
  Rev. B}\ }\textbf {\bibinfo {volume} {74}},\ \bibinfo {pages} {195312}
  (\bibinfo {year} {2006})}\BibitemShut {NoStop}%
\bibitem [{\citenamefont {König}\ \emph {et~al.}(2007)\citenamefont {König},
  \citenamefont {Wiedmann}, \citenamefont {Brüne}, \citenamefont {Roth},
  \citenamefont {Buhmann}, \citenamefont {Molenkamp}, \citenamefont {Qi},\ and\
  \citenamefont {Zhang}}]{science.1148047}%
  \BibitemOpen
  \bibfield  {author} {\bibinfo {author} {\bibfnamefont {Markus}\ \bibnamefont
  {König}}, \bibinfo {author} {\bibfnamefont {Steffen}\ \bibnamefont
  {Wiedmann}}, \bibinfo {author} {\bibfnamefont {Christoph}\ \bibnamefont
  {Brüne}}, \bibinfo {author} {\bibfnamefont {Andreas}\ \bibnamefont {Roth}},
  \bibinfo {author} {\bibfnamefont {Hartmut}\ \bibnamefont {Buhmann}}, \bibinfo
  {author} {\bibfnamefont {Laurens~W.}\ \bibnamefont {Molenkamp}}, \bibinfo
  {author} {\bibfnamefont {Xiao-Liang}\ \bibnamefont {Qi}}, \ and\ \bibinfo
  {author} {\bibfnamefont {Shou-Cheng}\ \bibnamefont {Zhang}},\ }\bibfield
  {title} {\enquote {\bibinfo {title} {Quantum spin hall insulator state in
  hgte quantum wells},}\ }\href {\doibase 10.1126/science.1148047} {\bibfield
  {journal} {\bibinfo  {journal} {Science}\ }\textbf {\bibinfo {volume}
  {318}},\ \bibinfo {pages} {766--770} (\bibinfo {year} {2007})}\BibitemShut
  {NoStop}%
\bibitem [{\citenamefont {Fu}\ and\ \citenamefont
  {Kane}(2007)}]{PhysRevB.76.045302}%
  \BibitemOpen
  \bibfield  {author} {\bibinfo {author} {\bibfnamefont {Liang}\ \bibnamefont
  {Fu}}\ and\ \bibinfo {author} {\bibfnamefont {C.~L.}\ \bibnamefont {Kane}},\
  }\bibfield  {title} {\enquote {\bibinfo {title} {Topological insulators with
  inversion symmetry},}\ }\href {\doibase 10.1103/PhysRevB.76.045302}
  {\bibfield  {journal} {\bibinfo  {journal} {Phys. Rev. B}\ }\textbf {\bibinfo
  {volume} {76}},\ \bibinfo {pages} {045302} (\bibinfo {year}
  {2007})}\BibitemShut {NoStop}%
\bibitem [{\citenamefont {Fu}\ \emph {et~al.}(2007)\citenamefont {Fu},
  \citenamefont {Kane},\ and\ \citenamefont {Mele}}]{PhysRevLett.98.106803}%
  \BibitemOpen
  \bibfield  {author} {\bibinfo {author} {\bibfnamefont {Liang}\ \bibnamefont
  {Fu}}, \bibinfo {author} {\bibfnamefont {C.~L.}\ \bibnamefont {Kane}}, \ and\
  \bibinfo {author} {\bibfnamefont {E.~J.}\ \bibnamefont {Mele}},\ }\bibfield
  {title} {\enquote {\bibinfo {title} {Topological insulators in three
  dimensions},}\ }\href {\doibase 10.1103/PhysRevLett.98.106803} {\bibfield
  {journal} {\bibinfo  {journal} {Phys. Rev. Lett.}\ }\textbf {\bibinfo
  {volume} {98}},\ \bibinfo {pages} {106803} (\bibinfo {year}
  {2007})}\BibitemShut {NoStop}%
\bibitem [{\citenamefont {Qi}\ \emph {et~al.}(2008)\citenamefont {Qi},
  \citenamefont {Hughes},\ and\ \citenamefont {Zhang}}]{PhysRevB.78.195424}%
  \BibitemOpen
  \bibfield  {author} {\bibinfo {author} {\bibfnamefont {Xiao-Liang}\
  \bibnamefont {Qi}}, \bibinfo {author} {\bibfnamefont {Taylor~L.}\
  \bibnamefont {Hughes}}, \ and\ \bibinfo {author} {\bibfnamefont {Shou-Cheng}\
  \bibnamefont {Zhang}},\ }\bibfield  {title} {\enquote {\bibinfo {title}
  {Topological field theory of time-reversal invariant insulators},}\ }\href
  {\doibase 10.1103/PhysRevB.78.195424} {\bibfield  {journal} {\bibinfo
  {journal} {Phys. Rev. B}\ }\textbf {\bibinfo {volume} {78}},\ \bibinfo
  {pages} {195424} (\bibinfo {year} {2008})}\BibitemShut {NoStop}%
\bibitem [{\citenamefont {Qi}\ and\ \citenamefont
  {Zhang}(2011)}]{RevModPhys.83.1057}%
  \BibitemOpen
  \bibfield  {author} {\bibinfo {author} {\bibfnamefont {Xiao-Liang}\
  \bibnamefont {Qi}}\ and\ \bibinfo {author} {\bibfnamefont {Shou-Cheng}\
  \bibnamefont {Zhang}},\ }\bibfield  {title} {\enquote {\bibinfo {title}
  {Topological insulators and superconductors},}\ }\href {\doibase
  10.1103/RevModPhys.83.1057} {\bibfield  {journal} {\bibinfo  {journal} {Rev.
  Mod. Phys.}\ }\textbf {\bibinfo {volume} {83}},\ \bibinfo {pages}
  {1057--1110} (\bibinfo {year} {2011})}\BibitemShut {NoStop}%
\bibitem [{\citenamefont {Chiu}\ \emph {et~al.}(2016)\citenamefont {Chiu},
  \citenamefont {Teo}, \citenamefont {Schnyder},\ and\ \citenamefont
  {Ryu}}]{RevModPhys.88.035005}%
  \BibitemOpen
  \bibfield  {author} {\bibinfo {author} {\bibfnamefont {Ching-Kai}\
  \bibnamefont {Chiu}}, \bibinfo {author} {\bibfnamefont {Jeffrey C.~Y.}\
  \bibnamefont {Teo}}, \bibinfo {author} {\bibfnamefont {Andreas~P.}\
  \bibnamefont {Schnyder}}, \ and\ \bibinfo {author} {\bibfnamefont {Shinsei}\
  \bibnamefont {Ryu}},\ }\bibfield  {title} {\enquote {\bibinfo {title}
  {Classification of topological quantum matter with symmetries},}\ }\href
  {\doibase 10.1103/RevModPhys.88.035005} {\bibfield  {journal} {\bibinfo
  {journal} {Rev. Mod. Phys.}\ }\textbf {\bibinfo {volume} {88}},\ \bibinfo
  {pages} {035005} (\bibinfo {year} {2016})}\BibitemShut {NoStop}%
\bibitem [{\citenamefont {Atiyah}\ \emph {et~al.}(1973)\citenamefont {Atiyah},
  \citenamefont {Patodi},\ and\ \citenamefont {Singer}}]{blms.5.2.229}%
  \BibitemOpen
  \bibfield  {author} {\bibinfo {author} {\bibfnamefont {M.~F.}\ \bibnamefont
  {Atiyah}}, \bibinfo {author} {\bibfnamefont {V.~K.}\ \bibnamefont {Patodi}},
  \ and\ \bibinfo {author} {\bibfnamefont {I.~M.}\ \bibnamefont {Singer}},\
  }\bibfield  {title} {\enquote {\bibinfo {title} {Spectral asymmetry and
  riemannian geometry},}\ }\href {\doibase
  https://doi.org/10.1112/blms/5.2.229} {\bibfield  {journal} {\bibinfo
  {journal} {Bulletin of the London Mathematical Society}\ }\textbf {\bibinfo
  {volume} {5}},\ \bibinfo {pages} {229--234} (\bibinfo {year}
  {1973})}\BibitemShut {NoStop}%
\bibitem [{\citenamefont {Weinberg}(1981)}]{PhysRevD.24.2669}%
  \BibitemOpen
  \bibfield  {author} {\bibinfo {author} {\bibfnamefont {Erick~J.}\
  \bibnamefont {Weinberg}},\ }\bibfield  {title} {\enquote {\bibinfo {title}
  {Index calculations for the fermion-vortex system},}\ }\href {\doibase
  10.1103/PhysRevD.24.2669} {\bibfield  {journal} {\bibinfo  {journal} {Phys.
  Rev. D}\ }\textbf {\bibinfo {volume} {24}},\ \bibinfo {pages} {2669--2673}
  (\bibinfo {year} {1981})}\BibitemShut {NoStop}%
\bibitem [{\citenamefont {Witten}(2016)}]{RevModPhys.88.035001}%
  \BibitemOpen
  \bibfield  {author} {\bibinfo {author} {\bibfnamefont {Edward}\ \bibnamefont
  {Witten}},\ }\bibfield  {title} {\enquote {\bibinfo {title} {Fermion path
  integrals and topological phases},}\ }\href {\doibase
  10.1103/RevModPhys.88.035001} {\bibfield  {journal} {\bibinfo  {journal}
  {Rev. Mod. Phys.}\ }\textbf {\bibinfo {volume} {88}},\ \bibinfo {pages}
  {035001} (\bibinfo {year} {2016})}\BibitemShut {NoStop}%
\bibitem [{\citenamefont {Kaplan}\ and\ \citenamefont
  {Sen}(2022)}]{PhysRevLett.128.251601}%
  \BibitemOpen
  \bibfield  {author} {\bibinfo {author} {\bibfnamefont {David~B.}\
  \bibnamefont {Kaplan}}\ and\ \bibinfo {author} {\bibfnamefont {Srimoyee}\
  \bibnamefont {Sen}},\ }\bibfield  {title} {\enquote {\bibinfo {title} {Index
  theorems, generalized hall currents, and topology for gapless defect
  fermions},}\ }\href {\doibase 10.1103/PhysRevLett.128.251601} {\bibfield
  {journal} {\bibinfo  {journal} {Phys. Rev. Lett.}\ }\textbf {\bibinfo
  {volume} {128}},\ \bibinfo {pages} {251601} (\bibinfo {year}
  {2022})}\BibitemShut {NoStop}%
\bibitem [{\citenamefont {Essin}\ and\ \citenamefont
  {Gurarie}(2011)}]{PhysRevB.84.125132}%
  \BibitemOpen
  \bibfield  {author} {\bibinfo {author} {\bibfnamefont {Andrew~M.}\
  \bibnamefont {Essin}}\ and\ \bibinfo {author} {\bibfnamefont {Victor}\
  \bibnamefont {Gurarie}},\ }\bibfield  {title} {\enquote {\bibinfo {title}
  {Bulk-boundary correspondence of topological insulators from their respective
  green's functions},}\ }\href {\doibase 10.1103/PhysRevB.84.125132} {\bibfield
   {journal} {\bibinfo  {journal} {Phys. Rev. B}\ }\textbf {\bibinfo {volume}
  {84}},\ \bibinfo {pages} {125132} (\bibinfo {year} {2011})}\BibitemShut
  {NoStop}%
\bibitem [{\citenamefont {Fidkowski}\ \emph {et~al.}(2011)\citenamefont
  {Fidkowski}, \citenamefont {Jackson},\ and\ \citenamefont
  {Klich}}]{PhysRevLett.107.036601}%
  \BibitemOpen
  \bibfield  {author} {\bibinfo {author} {\bibfnamefont {Lukasz}\ \bibnamefont
  {Fidkowski}}, \bibinfo {author} {\bibfnamefont {T.~S.}\ \bibnamefont
  {Jackson}}, \ and\ \bibinfo {author} {\bibfnamefont {Israel}\ \bibnamefont
  {Klich}},\ }\bibfield  {title} {\enquote {\bibinfo {title} {Model
  characterization of gapless edge modes of topological insulators using
  intermediate brillouin-zone functions},}\ }\href {\doibase
  10.1103/PhysRevLett.107.036601} {\bibfield  {journal} {\bibinfo  {journal}
  {Phys. Rev. Lett.}\ }\textbf {\bibinfo {volume} {107}},\ \bibinfo {pages}
  {036601} (\bibinfo {year} {2011})}\BibitemShut {NoStop}%
\bibitem [{\citenamefont {Fulga}\ \emph {et~al.}(2011)\citenamefont {Fulga},
  \citenamefont {Hassler}, \citenamefont {Akhmerov},\ and\ \citenamefont
  {Beenakker}}]{PhysRevB.83.155429}%
  \BibitemOpen
  \bibfield  {author} {\bibinfo {author} {\bibfnamefont {I.~C.}\ \bibnamefont
  {Fulga}}, \bibinfo {author} {\bibfnamefont {F.}~\bibnamefont {Hassler}},
  \bibinfo {author} {\bibfnamefont {A.~R.}\ \bibnamefont {Akhmerov}}, \ and\
  \bibinfo {author} {\bibfnamefont {C.~W.~J.}\ \bibnamefont {Beenakker}},\
  }\bibfield  {title} {\enquote {\bibinfo {title} {Scattering formula for the
  topological quantum number of a disordered multimode wire},}\ }\href
  {\doibase 10.1103/PhysRevB.83.155429} {\bibfield  {journal} {\bibinfo
  {journal} {Phys. Rev. B}\ }\textbf {\bibinfo {volume} {83}},\ \bibinfo
  {pages} {155429} (\bibinfo {year} {2011})}\BibitemShut {NoStop}%
\bibitem [{\citenamefont {Fulga}\ \emph {et~al.}(2012)\citenamefont {Fulga},
  \citenamefont {Hassler},\ and\ \citenamefont
  {Akhmerov}}]{PhysRevB.85.165409}%
  \BibitemOpen
  \bibfield  {author} {\bibinfo {author} {\bibfnamefont {I.~C.}\ \bibnamefont
  {Fulga}}, \bibinfo {author} {\bibfnamefont {F.}~\bibnamefont {Hassler}}, \
  and\ \bibinfo {author} {\bibfnamefont {A.~R.}\ \bibnamefont {Akhmerov}},\
  }\bibfield  {title} {\enquote {\bibinfo {title} {Scattering theory of
  topological insulators and superconductors},}\ }\href {\doibase
  10.1103/PhysRevB.85.165409} {\bibfield  {journal} {\bibinfo  {journal} {Phys.
  Rev. B}\ }\textbf {\bibinfo {volume} {85}},\ \bibinfo {pages} {165409}
  (\bibinfo {year} {2012})}\BibitemShut {NoStop}%
\bibitem [{\citenamefont {Peng}\ \emph {et~al.}(2017)\citenamefont {Peng},
  \citenamefont {Bao},\ and\ \citenamefont {von Oppen}}]{PhysRevB.95.235143}%
  \BibitemOpen
  \bibfield  {author} {\bibinfo {author} {\bibfnamefont {Yang}\ \bibnamefont
  {Peng}}, \bibinfo {author} {\bibfnamefont {Yimu}\ \bibnamefont {Bao}}, \ and\
  \bibinfo {author} {\bibfnamefont {Felix}\ \bibnamefont {von Oppen}},\
  }\bibfield  {title} {\enquote {\bibinfo {title} {Boundary green functions of
  topological insulators and superconductors},}\ }\href {\doibase
  10.1103/PhysRevB.95.235143} {\bibfield  {journal} {\bibinfo  {journal} {Phys.
  Rev. B}\ }\textbf {\bibinfo {volume} {95}},\ \bibinfo {pages} {235143}
  (\bibinfo {year} {2017})}\BibitemShut {NoStop}%
\bibitem [{\citenamefont {Yao}\ and\ \citenamefont {Wang}(2018)}]{yao2018}%
  \BibitemOpen
  \bibfield  {author} {\bibinfo {author} {\bibfnamefont {Shunyu}\ \bibnamefont
  {Yao}}\ and\ \bibinfo {author} {\bibfnamefont {Zhong}\ \bibnamefont {Wang}},\
  }\bibfield  {title} {\enquote {\bibinfo {title} {Edge states and topological
  invariants of non-hermitian systems},}\ }\href {\doibase
  10.1103/PhysRevLett.121.086803} {\bibfield  {journal} {\bibinfo  {journal}
  {Phys. Rev. Lett.}\ }\textbf {\bibinfo {volume} {121}},\ \bibinfo {pages}
  {086803} (\bibinfo {year} {2018})}\BibitemShut {NoStop}%
\bibitem [{\citenamefont {Yao}\ \emph {et~al.}(2018)\citenamefont {Yao},
  \citenamefont {Song},\ and\ \citenamefont {Wang}}]{yao20182}%
  \BibitemOpen
  \bibfield  {author} {\bibinfo {author} {\bibfnamefont {Shunyu}\ \bibnamefont
  {Yao}}, \bibinfo {author} {\bibfnamefont {Fei}\ \bibnamefont {Song}}, \ and\
  \bibinfo {author} {\bibfnamefont {Zhong}\ \bibnamefont {Wang}},\ }\bibfield
  {title} {\enquote {\bibinfo {title} {Non-hermitian chern bands},}\ }\href
  {\doibase 10.1103/PhysRevLett.121.136802} {\bibfield  {journal} {\bibinfo
  {journal} {Phys. Rev. Lett.}\ }\textbf {\bibinfo {volume} {121}},\ \bibinfo
  {pages} {136802} (\bibinfo {year} {2018})}\BibitemShut {NoStop}%
\bibitem [{\citenamefont {Bergholtz}\ \emph {et~al.}(2021)\citenamefont
  {Bergholtz}, \citenamefont {Budich},\ and\ \citenamefont
  {Kunst}}]{RevModPhys.93.015005}%
  \BibitemOpen
  \bibfield  {author} {\bibinfo {author} {\bibfnamefont {Emil~J.}\ \bibnamefont
  {Bergholtz}}, \bibinfo {author} {\bibfnamefont {Jan~Carl}\ \bibnamefont
  {Budich}}, \ and\ \bibinfo {author} {\bibfnamefont {Flore~K.}\ \bibnamefont
  {Kunst}},\ }\bibfield  {title} {\enquote {\bibinfo {title} {Exceptional
  topology of non-hermitian systems},}\ }\href {\doibase
  10.1103/RevModPhys.93.015005} {\bibfield  {journal} {\bibinfo  {journal}
  {Rev. Mod. Phys.}\ }\textbf {\bibinfo {volume} {93}},\ \bibinfo {pages}
  {015005} (\bibinfo {year} {2021})}\BibitemShut {NoStop}%
\bibitem [{\citenamefont {Kunst}\ \emph {et~al.}(2018)\citenamefont {Kunst},
  \citenamefont {Edvardsson}, \citenamefont {Budich},\ and\ \citenamefont
  {Bergholtz}}]{PhysRevLett.121.026808}%
  \BibitemOpen
  \bibfield  {author} {\bibinfo {author} {\bibfnamefont {Flore~K.}\
  \bibnamefont {Kunst}}, \bibinfo {author} {\bibfnamefont {Elisabet}\
  \bibnamefont {Edvardsson}}, \bibinfo {author} {\bibfnamefont {Jan~Carl}\
  \bibnamefont {Budich}}, \ and\ \bibinfo {author} {\bibfnamefont {Emil~J.}\
  \bibnamefont {Bergholtz}},\ }\bibfield  {title} {\enquote {\bibinfo {title}
  {Biorthogonal bulk-boundary correspondence in non-hermitian systems},}\
  }\href {\doibase 10.1103/PhysRevLett.121.026808} {\bibfield  {journal}
  {\bibinfo  {journal} {Phys. Rev. Lett.}\ }\textbf {\bibinfo {volume} {121}},\
  \bibinfo {pages} {026808} (\bibinfo {year} {2018})}\BibitemShut {NoStop}%
\bibitem [{\citenamefont {Xiong}(2018)}]{Xiong_2018}%
  \BibitemOpen
  \bibfield  {author} {\bibinfo {author} {\bibfnamefont {Ye}~\bibnamefont
  {Xiong}},\ }\bibfield  {title} {\enquote {\bibinfo {title} {Why does bulk
  boundary correspondence fail in some non-hermitian topological models},}\
  }\href {\doibase 10.1088/2399-6528/aab64a} {\bibfield  {journal} {\bibinfo
  {journal} {Journal of Physics Communications}\ }\textbf {\bibinfo {volume}
  {2}},\ \bibinfo {pages} {035043} (\bibinfo {year} {2018})}\BibitemShut
  {NoStop}%
\bibitem [{\citenamefont {Yang}\ \emph {et~al.}(2020)\citenamefont {Yang},
  \citenamefont {Zhang}, \citenamefont {Fang},\ and\ \citenamefont
  {Hu}}]{PhysRevLett.125.226402}%
  \BibitemOpen
  \bibfield  {author} {\bibinfo {author} {\bibfnamefont {Zhesen}\ \bibnamefont
  {Yang}}, \bibinfo {author} {\bibfnamefont {Kai}\ \bibnamefont {Zhang}},
  \bibinfo {author} {\bibfnamefont {Chen}\ \bibnamefont {Fang}}, \ and\
  \bibinfo {author} {\bibfnamefont {Jiangping}\ \bibnamefont {Hu}},\ }\bibfield
   {title} {\enquote {\bibinfo {title} {Non-hermitian bulk-boundary
  correspondence and auxiliary generalized brillouin zone theory},}\ }\href
  {\doibase 10.1103/PhysRevLett.125.226402} {\bibfield  {journal} {\bibinfo
  {journal} {Phys. Rev. Lett.}\ }\textbf {\bibinfo {volume} {125}},\ \bibinfo
  {pages} {226402} (\bibinfo {year} {2020})}\BibitemShut {NoStop}%
\bibitem [{\citenamefont {Song}\ \emph {et~al.}(2019)\citenamefont {Song},
  \citenamefont {Yao},\ and\ \citenamefont {Wang}}]{PhysRevLett.123.246801}%
  \BibitemOpen
  \bibfield  {author} {\bibinfo {author} {\bibfnamefont {Fei}\ \bibnamefont
  {Song}}, \bibinfo {author} {\bibfnamefont {Shunyu}\ \bibnamefont {Yao}}, \
  and\ \bibinfo {author} {\bibfnamefont {Zhong}\ \bibnamefont {Wang}},\
  }\bibfield  {title} {\enquote {\bibinfo {title} {Non-hermitian topological
  invariants in real space},}\ }\href {\doibase 10.1103/PhysRevLett.123.246801}
  {\bibfield  {journal} {\bibinfo  {journal} {Phys. Rev. Lett.}\ }\textbf
  {\bibinfo {volume} {123}},\ \bibinfo {pages} {246801} (\bibinfo {year}
  {2019})}\BibitemShut {NoStop}%
\bibitem [{\citenamefont {Lee}\ and\ \citenamefont
  {Thomale}(2019)}]{PhysRevB.99.201103}%
  \BibitemOpen
  \bibfield  {author} {\bibinfo {author} {\bibfnamefont {Ching~Hua}\
  \bibnamefont {Lee}}\ and\ \bibinfo {author} {\bibfnamefont {Ronny}\
  \bibnamefont {Thomale}},\ }\bibfield  {title} {\enquote {\bibinfo {title}
  {Anatomy of skin modes and topology in non-hermitian systems},}\ }\href
  {\doibase 10.1103/PhysRevB.99.201103} {\bibfield  {journal} {\bibinfo
  {journal} {Phys. Rev. B}\ }\textbf {\bibinfo {volume} {99}},\ \bibinfo
  {pages} {201103} (\bibinfo {year} {2019})}\BibitemShut {NoStop}%
\bibitem [{\citenamefont {Yokomizo}\ and\ \citenamefont
  {Murakami}(2019)}]{PhysRevLett.123.066404}%
  \BibitemOpen
  \bibfield  {author} {\bibinfo {author} {\bibfnamefont {Kazuki}\ \bibnamefont
  {Yokomizo}}\ and\ \bibinfo {author} {\bibfnamefont {Shuichi}\ \bibnamefont
  {Murakami}},\ }\bibfield  {title} {\enquote {\bibinfo {title} {Non-bloch band
  theory of non-hermitian systems},}\ }\href {\doibase
  10.1103/PhysRevLett.123.066404} {\bibfield  {journal} {\bibinfo  {journal}
  {Phys. Rev. Lett.}\ }\textbf {\bibinfo {volume} {123}},\ \bibinfo {pages}
  {066404} (\bibinfo {year} {2019})}\BibitemShut {NoStop}%
\bibitem [{\citenamefont {Regensburger}\ \emph {et~al.}(2012)\citenamefont
  {Regensburger}, \citenamefont {Bersch}, \citenamefont {Miri}, \citenamefont
  {Onishchukov}, \citenamefont {Christodoulides},\ and\ \citenamefont
  {Peschel}}]{Regensburger2012}%
  \BibitemOpen
  \bibfield  {author} {\bibinfo {author} {\bibfnamefont {Alois}\ \bibnamefont
  {Regensburger}}, \bibinfo {author} {\bibfnamefont {Christoph}\ \bibnamefont
  {Bersch}}, \bibinfo {author} {\bibfnamefont {Mohammad-Ali}\ \bibnamefont
  {Miri}}, \bibinfo {author} {\bibfnamefont {Georgy}\ \bibnamefont
  {Onishchukov}}, \bibinfo {author} {\bibfnamefont {Demetrios~N.}\ \bibnamefont
  {Christodoulides}}, \ and\ \bibinfo {author} {\bibfnamefont {Ulf}\
  \bibnamefont {Peschel}},\ }\bibfield  {title} {\enquote {\bibinfo {title}
  {Parity--time synthetic photonic lattices},}\ }\href {\doibase
  10.1038/nature11298} {\bibfield  {journal} {\bibinfo  {journal} {Nature}\
  }\textbf {\bibinfo {volume} {488}},\ \bibinfo {pages} {167--171} (\bibinfo
  {year} {2012})}\BibitemShut {NoStop}%
\bibitem [{\citenamefont {Rechtsman}\ \emph {et~al.}(2013)\citenamefont
  {Rechtsman}, \citenamefont {Zeuner}, \citenamefont {Plotnik}, \citenamefont
  {Lumer}, \citenamefont {Podolsky}, \citenamefont {Dreisow}, \citenamefont
  {Nolte}, \citenamefont {Segev},\ and\ \citenamefont
  {Szameit}}]{Rechtsman2013}%
  \BibitemOpen
  \bibfield  {author} {\bibinfo {author} {\bibfnamefont {Mikael~C.}\
  \bibnamefont {Rechtsman}}, \bibinfo {author} {\bibfnamefont {Julia~M.}\
  \bibnamefont {Zeuner}}, \bibinfo {author} {\bibfnamefont {Yonatan}\
  \bibnamefont {Plotnik}}, \bibinfo {author} {\bibfnamefont {Yaakov}\
  \bibnamefont {Lumer}}, \bibinfo {author} {\bibfnamefont {Daniel}\
  \bibnamefont {Podolsky}}, \bibinfo {author} {\bibfnamefont {Felix}\
  \bibnamefont {Dreisow}}, \bibinfo {author} {\bibfnamefont {Stefan}\
  \bibnamefont {Nolte}}, \bibinfo {author} {\bibfnamefont {Mordechai}\
  \bibnamefont {Segev}}, \ and\ \bibinfo {author} {\bibfnamefont {Alexander}\
  \bibnamefont {Szameit}},\ }\bibfield  {title} {\enquote {\bibinfo {title}
  {Photonic floquet topological insulators},}\ }\href {\doibase
  10.1038/nature12066} {\bibfield  {journal} {\bibinfo  {journal} {Nature}\
  }\textbf {\bibinfo {volume} {496}},\ \bibinfo {pages} {196--200} (\bibinfo
  {year} {2013})}\BibitemShut {NoStop}%
\bibitem [{\citenamefont {Weimann}\ \emph {et~al.}(2017)\citenamefont
  {Weimann}, \citenamefont {Kremer}, \citenamefont {Plotnik}, \citenamefont
  {Lumer}, \citenamefont {Nolte}, \citenamefont {Makris}, \citenamefont
  {Segev}, \citenamefont {Rechtsman},\ and\ \citenamefont
  {Szameit}}]{Weimann2017}%
  \BibitemOpen
  \bibfield  {author} {\bibinfo {author} {\bibfnamefont {S.}~\bibnamefont
  {Weimann}}, \bibinfo {author} {\bibfnamefont {M.}~\bibnamefont {Kremer}},
  \bibinfo {author} {\bibfnamefont {Y.}~\bibnamefont {Plotnik}}, \bibinfo
  {author} {\bibfnamefont {Y.}~\bibnamefont {Lumer}}, \bibinfo {author}
  {\bibfnamefont {S.}~\bibnamefont {Nolte}}, \bibinfo {author} {\bibfnamefont
  {K.~G.}\ \bibnamefont {Makris}}, \bibinfo {author} {\bibfnamefont
  {M.}~\bibnamefont {Segev}}, \bibinfo {author} {\bibfnamefont {M.~C.}\
  \bibnamefont {Rechtsman}}, \ and\ \bibinfo {author} {\bibfnamefont
  {A.}~\bibnamefont {Szameit}},\ }\bibfield  {title} {\enquote {\bibinfo
  {title} {Topologically protected bound states in photonic
  parity--time-symmetric crystals},}\ }\href {\doibase 10.1038/nmat4811}
  {\bibfield  {journal} {\bibinfo  {journal} {Nature Materials}\ }\textbf
  {\bibinfo {volume} {16}},\ \bibinfo {pages} {433--438} (\bibinfo {year}
  {2017})}\BibitemShut {NoStop}%
\bibitem [{\citenamefont {Ashida}\ \emph {et~al.}(2020)\citenamefont {Ashida},
  \citenamefont {Gong},\ and\ \citenamefont
  {Ueda}}]{doi:10.1080/00018732.2021.1876991}%
  \BibitemOpen
  \bibfield  {author} {\bibinfo {author} {\bibfnamefont {Yuto}\ \bibnamefont
  {Ashida}}, \bibinfo {author} {\bibfnamefont {Zongping}\ \bibnamefont {Gong}},
  \ and\ \bibinfo {author} {\bibfnamefont {Masahito}\ \bibnamefont {Ueda}},\
  }\bibfield  {title} {\enquote {\bibinfo {title} {Non-hermitian physics},}\
  }\href {\doibase 10.1080/00018732.2021.1876991} {\bibfield  {journal}
  {\bibinfo  {journal} {Advances in Physics}\ }\textbf {\bibinfo {volume}
  {69}},\ \bibinfo {pages} {249--435} (\bibinfo {year} {2020})},\ \Eprint
  {http://arxiv.org/abs/https://doi.org/10.1080/00018732.2021.1876991}
  {https://doi.org/10.1080/00018732.2021.1876991} \BibitemShut {NoStop}%
\bibitem [{\citenamefont {Beenakker}(1997)}]{RevModPhys.69.731}%
  \BibitemOpen
  \bibfield  {author} {\bibinfo {author} {\bibfnamefont {C.~W.~J.}\
  \bibnamefont {Beenakker}},\ }\bibfield  {title} {\enquote {\bibinfo {title}
  {Random-matrix theory of quantum transport},}\ }\href {\doibase
  10.1103/RevModPhys.69.731} {\bibfield  {journal} {\bibinfo  {journal} {Rev.
  Mod. Phys.}\ }\textbf {\bibinfo {volume} {69}},\ \bibinfo {pages} {731--808}
  (\bibinfo {year} {1997})}\BibitemShut {NoStop}%
\bibitem [{\citenamefont {Peng}\ \emph {et~al.}(2014)\citenamefont {Peng},
  \citenamefont {{\"O}zdemir}, \citenamefont {Rotter}, \citenamefont {Yilmaz},
  \citenamefont {Liertzer}, \citenamefont {Monifi}, \citenamefont {Bender},
  \citenamefont {Nori},\ and\ \citenamefont {Yang}}]{peng2014loss}%
  \BibitemOpen
  \bibfield  {author} {\bibinfo {author} {\bibfnamefont {B}~\bibnamefont
  {Peng}}, \bibinfo {author} {\bibfnamefont {{\c{S}}K}~\bibnamefont
  {{\"O}zdemir}}, \bibinfo {author} {\bibfnamefont {S}~\bibnamefont {Rotter}},
  \bibinfo {author} {\bibfnamefont {H}~\bibnamefont {Yilmaz}}, \bibinfo
  {author} {\bibfnamefont {M}~\bibnamefont {Liertzer}}, \bibinfo {author}
  {\bibfnamefont {F}~\bibnamefont {Monifi}}, \bibinfo {author} {\bibfnamefont
  {CM}~\bibnamefont {Bender}}, \bibinfo {author} {\bibfnamefont
  {F}~\bibnamefont {Nori}}, \ and\ \bibinfo {author} {\bibfnamefont
  {L}~\bibnamefont {Yang}},\ }\bibfield  {title} {\enquote {\bibinfo {title}
  {Loss-induced suppression and revival of lasing},}\ }\href@noop {} {\bibfield
   {journal} {\bibinfo  {journal} {Science}\ }\textbf {\bibinfo {volume}
  {346}},\ \bibinfo {pages} {328--332} (\bibinfo {year} {2014})}\BibitemShut
  {NoStop}%
\bibitem [{\citenamefont {Longhi}\ \emph {et~al.}(2015)\citenamefont {Longhi},
  \citenamefont {Gatti},\ and\ \citenamefont {Valle}}]{Longhi2015}%
  \BibitemOpen
  \bibfield  {author} {\bibinfo {author} {\bibfnamefont {Stefano}\ \bibnamefont
  {Longhi}}, \bibinfo {author} {\bibfnamefont {Davide}\ \bibnamefont {Gatti}},
  \ and\ \bibinfo {author} {\bibfnamefont {Giuseppe~Della}\ \bibnamefont
  {Valle}},\ }\bibfield  {title} {\enquote {\bibinfo {title} {Robust light
  transport in non-hermitian photonic lattices},}\ }\href {\doibase
  10.1038/srep13376} {\bibfield  {journal} {\bibinfo  {journal} {Scientific
  Reports}\ }\textbf {\bibinfo {volume} {5}},\ \bibinfo {pages} {13376}
  (\bibinfo {year} {2015})}\BibitemShut {NoStop}%
\bibitem [{\citenamefont {Zhao}\ \emph {et~al.}(2019)\citenamefont {Zhao},
  \citenamefont {Qiao}, \citenamefont {Wu}, \citenamefont {Midya},
  \citenamefont {Longhi},\ and\ \citenamefont
  {Feng}}]{doi:10.1126/science.aay1064}%
  \BibitemOpen
  \bibfield  {author} {\bibinfo {author} {\bibfnamefont {Han}\ \bibnamefont
  {Zhao}}, \bibinfo {author} {\bibfnamefont {Xingdu}\ \bibnamefont {Qiao}},
  \bibinfo {author} {\bibfnamefont {Tianwei}\ \bibnamefont {Wu}}, \bibinfo
  {author} {\bibfnamefont {Bikashkali}\ \bibnamefont {Midya}}, \bibinfo
  {author} {\bibfnamefont {Stefano}\ \bibnamefont {Longhi}}, \ and\ \bibinfo
  {author} {\bibfnamefont {Liang}\ \bibnamefont {Feng}},\ }\bibfield  {title}
  {\enquote {\bibinfo {title} {Non-hermitian topological light steering},}\
  }\href {\doibase 10.1126/science.aay1064} {\bibfield  {journal} {\bibinfo
  {journal} {Science}\ }\textbf {\bibinfo {volume} {365}},\ \bibinfo {pages}
  {1163--1166} (\bibinfo {year} {2019})},\ \Eprint
  {http://arxiv.org/abs/https://www.science.org/doi/pdf/10.1126/science.aay1064}
  {https://www.science.org/doi/pdf/10.1126/science.aay1064} \BibitemShut
  {NoStop}%
\bibitem [{Note1()}]{Note1}%
  \BibitemOpen
  \bibinfo {note} {Another difference compared to the Hermitian cases is
  $\protect \qopname \relax m{lim}_{\omega \rightarrow 0} \Gamma r(\omega )$ is
  not necessarily Hermitian \cite {PhysRevB.85.165409}}\BibitemShut {NoStop}%
\bibitem [{\citenamefont {Kawabata}\ \emph {et~al.}(2019)\citenamefont
  {Kawabata}, \citenamefont {Shiozaki}, \citenamefont {Ueda},\ and\
  \citenamefont {Sato}}]{PhysRevX.9.041015}%
  \BibitemOpen
  \bibfield  {author} {\bibinfo {author} {\bibfnamefont {Kohei}\ \bibnamefont
  {Kawabata}}, \bibinfo {author} {\bibfnamefont {Ken}\ \bibnamefont
  {Shiozaki}}, \bibinfo {author} {\bibfnamefont {Masahito}\ \bibnamefont
  {Ueda}}, \ and\ \bibinfo {author} {\bibfnamefont {Masatoshi}\ \bibnamefont
  {Sato}},\ }\bibfield  {title} {\enquote {\bibinfo {title} {Symmetry and
  topology in non-hermitian physics},}\ }\href {\doibase
  10.1103/PhysRevX.9.041015} {\bibfield  {journal} {\bibinfo  {journal} {Phys.
  Rev. X}\ }\textbf {\bibinfo {volume} {9}},\ \bibinfo {pages} {041015}
  (\bibinfo {year} {2019})}\BibitemShut {NoStop}%
\bibitem [{\citenamefont {Kawabata}\ \emph {et~al.}(2020)\citenamefont
  {Kawabata}, \citenamefont {Okuma},\ and\ \citenamefont
  {Sato}}]{PhysRevB.101.195147}%
  \BibitemOpen
  \bibfield  {author} {\bibinfo {author} {\bibfnamefont {Kohei}\ \bibnamefont
  {Kawabata}}, \bibinfo {author} {\bibfnamefont {Nobuyuki}\ \bibnamefont
  {Okuma}}, \ and\ \bibinfo {author} {\bibfnamefont {Masatoshi}\ \bibnamefont
  {Sato}},\ }\bibfield  {title} {\enquote {\bibinfo {title} {Non-bloch band
  theory of non-hermitian hamiltonians in the symplectic class},}\ }\href
  {\doibase 10.1103/PhysRevB.101.195147} {\bibfield  {journal} {\bibinfo
  {journal} {Phys. Rev. B}\ }\textbf {\bibinfo {volume} {101}},\ \bibinfo
  {pages} {195147} (\bibinfo {year} {2020})}\BibitemShut {NoStop}%
\bibitem [{\citenamefont {Groth}\ \emph {et~al.}(2014)\citenamefont {Groth},
  \citenamefont {Wimmer}, \citenamefont {Akhmerov},\ and\ \citenamefont
  {Waintal}}]{Groth_2014}%
  \BibitemOpen
  \bibfield  {author} {\bibinfo {author} {\bibfnamefont {Christoph~W}\
  \bibnamefont {Groth}}, \bibinfo {author} {\bibfnamefont {Michael}\
  \bibnamefont {Wimmer}}, \bibinfo {author} {\bibfnamefont {Anton~R}\
  \bibnamefont {Akhmerov}}, \ and\ \bibinfo {author} {\bibfnamefont {Xavier}\
  \bibnamefont {Waintal}},\ }\bibfield  {title} {\enquote {\bibinfo {title}
  {Kwant: a software package for quantum transport},}\ }\href {\doibase
  10.1088/1367-2630/16/6/063065} {\bibfield  {journal} {\bibinfo  {journal}
  {New Journal of Physics}\ }\textbf {\bibinfo {volume} {16}},\ \bibinfo
  {pages} {063065} (\bibinfo {year} {2014})}\BibitemShut {NoStop}%
\bibitem [{\citenamefont {Xue}\ \emph {et~al.}(2021)\citenamefont {Xue},
  \citenamefont {Li}, \citenamefont {Hu}, \citenamefont {Song},\ and\
  \citenamefont {Wang}}]{PhysRevB.103.L241408}%
  \BibitemOpen
  \bibfield  {author} {\bibinfo {author} {\bibfnamefont {Wen-Tan}\ \bibnamefont
  {Xue}}, \bibinfo {author} {\bibfnamefont {Ming-Rui}\ \bibnamefont {Li}},
  \bibinfo {author} {\bibfnamefont {Yu-Min}\ \bibnamefont {Hu}}, \bibinfo
  {author} {\bibfnamefont {Fei}\ \bibnamefont {Song}}, \ and\ \bibinfo {author}
  {\bibfnamefont {Zhong}\ \bibnamefont {Wang}},\ }\bibfield  {title} {\enquote
  {\bibinfo {title} {Simple formulas of directional amplification from
  non-bloch band theory},}\ }\href {\doibase 10.1103/PhysRevB.103.L241408}
  {\bibfield  {journal} {\bibinfo  {journal} {Phys. Rev. B}\ }\textbf {\bibinfo
  {volume} {103}},\ \bibinfo {pages} {L241408} (\bibinfo {year}
  {2021})}\BibitemShut {NoStop}%
\bibitem [{\citenamefont {Li}\ and\ \citenamefont
  {Wan}(2022)}]{PhysRevB.105.045122}%
  \BibitemOpen
  \bibfield  {author} {\bibinfo {author} {\bibfnamefont {Haoshu}\ \bibnamefont
  {Li}}\ and\ \bibinfo {author} {\bibfnamefont {Shaolong}\ \bibnamefont
  {Wan}},\ }\bibfield  {title} {\enquote {\bibinfo {title} {Exact formulas of
  the end-to-end green's functions in non-hermitian systems},}\ }\href
  {\doibase 10.1103/PhysRevB.105.045122} {\bibfield  {journal} {\bibinfo
  {journal} {Phys. Rev. B}\ }\textbf {\bibinfo {volume} {105}},\ \bibinfo
  {pages} {045122} (\bibinfo {year} {2022})}\BibitemShut {NoStop}%
\end{thebibliography}%

\end{document}